\documentclass[twocolumn]{aastex631}

\usepackage{amsmath}	
\usepackage{amssymb}	
\usepackage{wasysym}
\usepackage{multirow}
\usepackage[flushleft]{threeparttable}
\usepackage{booktabs}

\shorttitle{TESS all-sky oscillating giants}
\shortauthors{Hon et al.}
\graphicspath{{./}{figures/}}

\begin{document}
\title{A `Quick Look' at All-Sky Galactic Archeology with TESS:\\ 158,000 Oscillating Red Giants from the MIT Quick-Look Pipeline}

\correspondingauthor{Marc Hon}
\email{mtyhon@hawaii.edu}

\author[0000-0003-2400-6960]{Marc Hon}
\altaffiliation{NASA Hubble Fellow}
\affiliation{Institute for Astronomy, University of Hawai`i,
2680 Woodlawn Drive, Honolulu, HI 96822, USA}
\affiliation{School of Physics, The University of New South Wales,
Sydney NSW 2052, Australia}

\author{Daniel Huber}
\affiliation{Institute for Astronomy, University of Hawai`i,
2680 Woodlawn Drive, Honolulu, HI 96822, USA}

\author{James S. Kuszlewicz}
\affiliation{Landessternwarte, Zentrum f\"{u}r Astronomie der Universit\"{a}t Heidelberg, K\"{o}nigstuhl 12, 69117, Heidelberg, Germany}

\author{Dennis Stello}
\affiliation{School of Physics, The University of New South Wales,
Sydney NSW 2052, Australia}
\affiliation{Stellar Astrophysics Centre, Department of Physics and Astronomy, Aarhus University,
Ny Munkegade 120, DK-8000 Aarhus C, Denmark}
\affiliation{Sydney Institute for Astronomy (SIfA), School of Physics, University of Sydney,
NSW 2006, Australia}
\affiliation{ARC Centre of Excellence for Astrophysics in Three Dimensions (ASTRO-3D), Australia}

\author{Sanjib Sharma}
\affiliation{Sydney Institute for Astronomy (SIfA), School of Physics, University of Sydney,
NSW 2006, Australia}
\affiliation{ARC Centre of Excellence for Astrophysics in Three Dimensions (ASTRO-3D), Australia}

\author{Jamie Tayar}
\altaffiliation{NASA Hubble Fellow}
\affiliation{Institute for Astronomy, University of Hawai`i, 2680 Woodlawn Drive, Honolulu, HI 96822, USA}

\author{Joel C. Zinn}
\altaffiliation{NSF Astronomy and Astrophysics Postdoctoral Fellow.}
\affiliation{Department of Astrophysics, American Museum of Natural History, Central Park West at 79th Street, New York, NY 10024, USA}
\affiliation{School of Physics, The University of New South Wales,
Sydney NSW 2052, Australia}

\author{Mathieu Vrard}
\affiliation{Department of Astronomy, The Ohio State University, Columbus, OH 43210, USA}

\author{Marc H. Pinsonneault}
\affiliation{Department of Astronomy, The Ohio State University, Columbus, OH 43210, USA}

\begin{abstract}

We present the first near all-sky yield of oscillating red giants from the prime mission data of NASA's Transiting Exoplanet Survey Satellite (TESS). We apply machine learning towards long-cadence TESS photometry from the first data release by the MIT Quick-Look Pipeline to automatically detect the presence of red giant oscillations in frequency power spectra. The detected targets are conservatively vetted to produce a total of 158,505 oscillating red giants, which is an order of magnitude increase over the yield from \textit{Kepler} and K2 and a lower limit to the possible yield of oscillating giants across TESS's nominal mission. For each detected target, we report effective temperatures and radii derived from colors and \textit{Gaia} parallaxes, as well as estimates of their frequency at maximum oscillation power. Using our measurements, we present the first near all-sky \textit{Gaia}-asteroseismology mass map, which shows global structures consistent with the expected stellar populations of our Galaxy.
To demonstrate the strong potential of TESS asteroseismology for Galactic archeology even with only one month of observations, we identify 354 new candidates for oscillating giants in the Galactic halo, display the vertical mass gradient of the Milky Way disk, and visualize correlations of stellar masses with kinematic phase space substructures, velocity dispersions, and $\alpha$-abundances.

\end{abstract}



\section{Introduction} \label{sec:intro}

With high-precision, uninterrupted photometry from previous space-borne missions CoRoT \citep{Baglin_2006}, \textit{Kepler} \citep{Borucki_2010} and K2 \citep{Howell_2014}, red giant asteroseismology has emerged as a powerful tool for probing stellar populations around the Milky Way. The precise measurements of masses, radii, and ages across thousands of field red giants offered by asteroseismology has brought forth new avenues for detailed studies of Galactic archeology whereby such measurements are combined with kinematic and spectroscopic information to build a map describing the chemical and dynamical evolution of the Milky Way (e.g., \citealt{Miglio_2013,Casagrande_2016, Aguirre_2018, Sharma_2019, Miglio_2021}).

However, efforts to apply asteroseismology across the Galaxy have been mainly hampered by the size of the observing windows provided by previous space missions. CoRoT observed a number of 4 square degree fields focused towards two opposite directions and observed $\sim$$10^6$ red giants \citep{Baglin_2016}, for which $\sim$3,000 thus far have had seismic measurements \citep{Peralta_2018}. Meanwhile, \textit{Kepler} observed one region in the northern hemisphere spanning 100 square degrees, which resulted in $\sim$20,000 detections \citep{Hon_2019}. The K2 mission observed 18 `\textit{Kepler} field'-sized regions along the ecliptic, motivating the establishment of the K2 Galactic archeology Program \citep{Stello_2017, Zinn_2020}, which will further increase the number of detections by $\sim$30,000 (Zinn et al., in review).
Now, we seek to extend the scope of red giant asteroseismology towards the entire sky with NASA's Transiting Exoplanet Survey Satellite \citep[TESS]{Ricker_2014}. Throughout the first two years of its nominal mission, TESS surveyed nearly the entire sky by observing 13 distinct sectors in both northern and southern hemispheres. Each sector covers a 96$^{\circ}$x~24$^{\circ}$ field of view and yields Full Frame Images (FFIs) that provide high-precision flux measurements of millions of stars for up to 27 days at a 30-minute cadence. Meanwhile, stars in overlapping sectors are observed for at least 27 days, with longer observational duration typically for stars closer to the ecliptic poles where there is maximal overlap between sectors.
This unprecedented level of sky coverage is expected to increase the number of red giants with detected oscillations by about an order of magnitude over its predecessors.

A big challenge faced by asteroseismology with TESS is the survey's observational duration. The majority of targets from the first two years of TESS observations will only be observed for 27 days, resulting in lower frequency resolution compared to \textit{Kepler} and K2. However, the asteroseismic inference of stellar populations using TESS data is still viable. Depending on the availability of precise seismic measurements, \citet{Aguirre_2020} showed that it is possible to obtain precision levels of up to $\sim$3\% for radii, $\sim$5\% for mass, and $\sim$20\% for ages when incorporating both spectroscopy and parallax information in modelling the brightest ($V\sim6-7$) TESS red giants. Similar precision levels were found by \citet{Stello_2021} when comparing seismic results from TESS data against those from 4-year \textit{Kepler} data for $\sim$2,000 red giants observed by both \textit{Kepler} and TESS. Furthermore, \citet{Mackereth_2021} recovered median mass uncertainties of $\sim8$\% and age uncertainties of $\sim26$\% for $\sim$1,700 red giants with $G < 11$ near the southern ecliptic pole, where FFI targets were observed by TESS for up to a year. \citet{Mackereth_2021} additionally indicate that the seismic ages of giants from TESS are sufficiently precise to distinguish chemically- and kinematically-defined structures within the Milky Way disk. Although the current outlook for red giant asteroseismology with TESS is promising, it has yet to be extended to the majority of TESS FFI targets. Not only
is this because significantly more effort is required to systematically detect populations of oscillating red giants across the millions of TESS FFI targets, but detailed seismic parameters like the large frequency separation are difficult to measure with one-month long observations \citep{Hekker_2012, Mosser_2019, Mackereth_2021, Stello_2021}. The frequency at maximum oscillation power, $\nu_{\mathrm{max}}$, however, can be measured from any star with a detected oscillation power excess, making it a more accessible seismic measurement for inferring fundamental stellar properties.


The systematic detection of red giant oscillations, which has previously posed a significant challenge in the field, can now be facilitated using machine learning algorithms \citep{Hon_2018, Hon_2019}. Such tools are vital in the era of TESS given the enormous volume of data the mission will provide, and indeed particular emphasis on the analyses of TESS-like datasets in asteroseismic machine learning has been provided in recent work (e.g., \citealt{Armstrong_2016, Hon_2018b, Bugnet_2019, Kgoadi_2019, Kuszlewicz_2020, Audenaert_2021}). However, an important aspect of supervised learning \textemdash{} which comprise most of current machine learning algorithms in asteroseismology \textemdash{} is the presence of training data that is expected to be representative of the data upon which inference is to be performed. In comparison to \textit{Kepler} and K2, TESS has photometric qualities and instrumental noise sources that are unique to its observing instrument. This property makes it challenging to generalize algorithms that learn from \textit{Kepler}/K2 data towards TESS data. 


\begin{figure}
    \centering
	\includegraphics[width=1.\columnwidth]{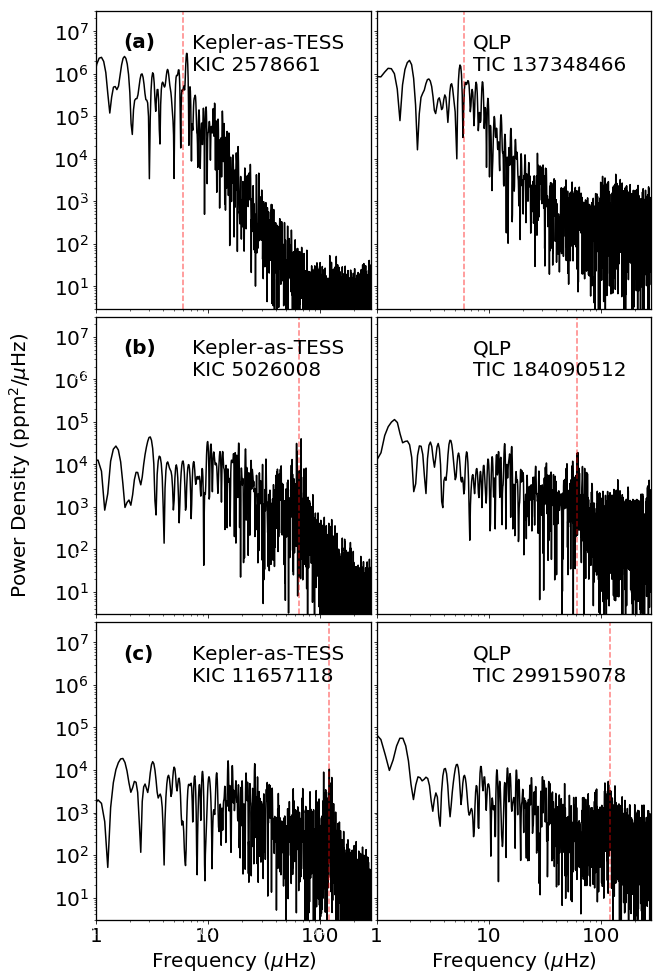}
    \caption{A comparison of red giant frequency power spectra from \textit{Kepler} and TESS. Each row corresponds to a red giant oscillating at a specific $\nu_{\mathrm{max}}$ value (red line). Power spectra in the left column correspond to 27-day light curves from \textit{Kepler}, while those in the right column correspond to the same \textit{Kepler} targets but observed by TESS. }
    \label{fig:1}
\end{figure}

Despite these challenges, we proceed with applying machine learning to perform the first all-sky detection of red giants across the full two-year TESS Primary Mission. We examine all light curves that have been extracted from FFIs by the MIT Quick-Look Pipeline \citep[QLP]{Huang_2020, Huang_2020b} as High Level Science Products. In conjunction with detecting a substantial number of oscillating TESS red giants, we also provide a `first look' at TESS's asteroseismic potential for full-sky Galactic archeology.

\section{Data}\label{sec:data}
We use all FFI light curves from the QLP team's first data release\footnote{\href{https://archive.stsci.edu/hlsp/qlp/}{https://archive.stsci.edu/hlsp/qlp/}}, which comprises observations across Sectors 1-26. This data release includes all targets brighter than a TESS magnitude of 13.5 and contains 24,376,080 light curves that have an observing cadence of $\sim$30 minutes. In this study, we use only the `raw' light curves obtained using simple aperture photometry~{\fontfamily{qcr}\selectfont
(SAP\_FLUX)}. We do not use the
detrended light curves {\fontfamily{qcr}\selectfont
(KSPSAP\_FLUX)}, which are optimized for planet searches and therefore have stellar variability over long timescales filtered out. Additionally, we only use timestamps with a good quality flag {\fontfamily{qcr}\selectfont
(QUALITY=0)}. Next, we high-pass filter each light curve using a boxcar filter with a width of 2-days and compute the light curve's power spectral density using the generalized Lomb-Scargle periodogram \citep{Lomb_1976, Scargle_1982}. During this processing, we exclude light curves which have most ($\geq95\%$) of their timestamps empty, resulting in a total of 23,962,744 power spectra corresponding to 14,702,113 unique stars observed by TESS\footnote{A fraction of targets are observed in multiple TESS sectors within overlapping fields of view, resulting in more power spectra than unique stars within the data release.}.

\section{Method} \label{sec:methods}

\subsection{Neural Networks}
We use the deep learning method as described by \citet{Hon_2018} to `visually' detect the presence of red giant oscillations within frequency power spectra. In particular, two convolutional neural networks are applied towards 2D images of log-log\footnote{The power spectra are represented in log-frequency on the x-axis and log-power density on the y-axis.} power spectra. By binning each star's power spectrum into a 128x128 binary image, the first neural network outputs a classification score, $p_{\mathrm{det}}$, between 0 and 1 indicating the likelihood that red giant oscillations are visible within the spectrum. The second network, which trained independently from the first, is a regression network that measures the frequency at maximum oscillation power, $\nu_{\mathrm{max}}$ and its corresponding uncertainty, $\sigma_{\nu_{\mathrm{max}}}$. Specific details of the structure of both networks are provided in Appendix \ref{Appendix:Network_Architecture}.

\subsection{Training Data}
\label{sec:training_data}
To train and test the classifier, we use KEPSEISMIC\footnote{\href{https://archive.stsci.edu/prepds/kepseismic/}{https://archive.stsci.edu/prepds/kepseismic/}} light curves \citep{KEPSEISMIC} of 196,581 \textit{Kepler} targets of which 21,914 were identified to be oscillating red giants from the detections provided by \citet{Hon_2019}. For the $\nu_{\mathrm{max}}$ regression network, we use a subset of the oscillating giants, specifically 16,194 red giants that have $\nu_{\mathrm{max}}$ values measured from the asteroseismic data pipeline by \citet{Yu_2018}. For both classification and regression datasets, we allocate 70\% of data to be used for training, 15\% to be used to internally validate and tune the network, and the remaining 15\% for network testing, which we describe in Section \ref{sec:threshold}.

Because we want to train the deep learning method on light curves of comparable duration to TESS data, each 4-year light curve in the training sets is segmented into multiple, unique 27-day segments. Due to the stochastic behaviour of solar-like oscillations, there is, however, no guarantee that oscillations detectable in a 4-year light curve will be visible over its constituent 27-day light curve segments. Using 4-year 
$\nu_{\mathrm{max}}$ measurements from \citet{Yu_2018}, we apply the formalism of \citet{Chaplin_2011} to statistically assess the significance of the oscillations within $\pm0.66~\nu_{\mathrm{max}}^{0.88}$ around $\nu_{\mathrm{max}}$ for the power spectrum of each 27-day light curve segment\footnote{Following \citet{Mosser_2012b}, this frequency range is approximately twice the full width at half maximum of the oscillation power envelope.}. For a power spectrum of a light curve segment to be included in the training set, it must have an oscillation SNR exceeding the 1\% false alarm threshold of a pure white noise spectrum. Because the $\nu_{\mathrm{max}}$ for each power spectrum is already known from 4-year \textit{Kepler} data, we measure the oscillation SNR directly from the spectrum rather than estimating it from fundamental stellar properties as per the original \citet{Chaplin_2011} approach. Our final classification training set comprising \textit{Kepler} data contains 714,820 light curves showing red giant oscillations and 4,959,892 light curves without red giant oscillations. Meanwhile, there are a total of 550,786 red giant power spectra for training the $\nu_{\mathrm{max}}$ regression network.


\begin{figure}
    \centering
	\includegraphics[width=1.\columnwidth]{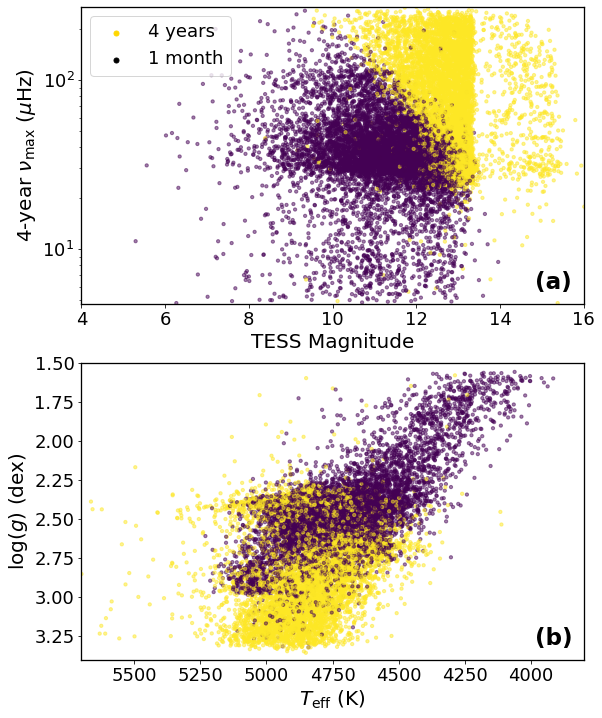}
    \caption{The estimated difference in yield for oscillating \textit{Kepler} red giants between 4-year \textit{Kepler} data and 1-month TESS data, depicted with a $\nu_{\mathrm{max}}$-magnitude diagram in panel (a) and a Kiel diagram in panel (b). 
    The $\nu_{\mathrm{max}}$ and $\log(g)$ values shown here are from \citet{Yu_2018}, while the estimate for the 1-month yield is calculated using the \citet{Chaplin_2011} formalism for predicting seismic yields adapted to a 27-day observations following \citet{Campante_2016, Schofield_2019}.}
    \label{fig:training_data}
\end{figure}


One major challenge with using our \textit{Kepler} training set is its difference in photometric data quality compared to TESS. In particular, TESS observations have lower photometric precision levels and are more susceptible to crowding effects \citep{Sullivan_2015}. Additionally, TESS data may contain poorly characterized systematic noise. Power spectra from TESS data are therefore expected to have greater noise levels compared to \textit{Kepler} data that we use for training, as demonstrated in Figure \ref{fig:1}. As a result, the visibility of the oscillations for low luminosity (high $\nu_{\mathrm{max}}$) red giants, particularly for those at the red giant branch, are diminished when approaching fainter TESS magnitudes as shown in Figure Figure \ref{fig:training_data}a. Here, it can be seen that for TESS magnitudes $\apprge10$, high $\nu_{\mathrm{max}}$ oscillations that can be seen with 4 years of data are not expected to be observed in 1 month of data, in agreement with the analytical predictions made by \citet{Mosser_2019}. Consequently, we expect to detect oscillations primarily in giants with $\log(g)\apprge3.0$ dex as shown in Figure \ref{fig:training_data}b.

Besides the expected yield, the difference in photometric data quality between \textit{Kepler} and TESS will also affect how well our networks can perform; it is thus important to identify this difference in performance so that we can determine a decision threshold that can adequately recover oscillating giants from TESS data.
\\
\\

\subsection{Setting a Network Decision Threshold}\label{sec:threshold}

\begin{figure}
    \centering
	\includegraphics[width=1.\columnwidth]{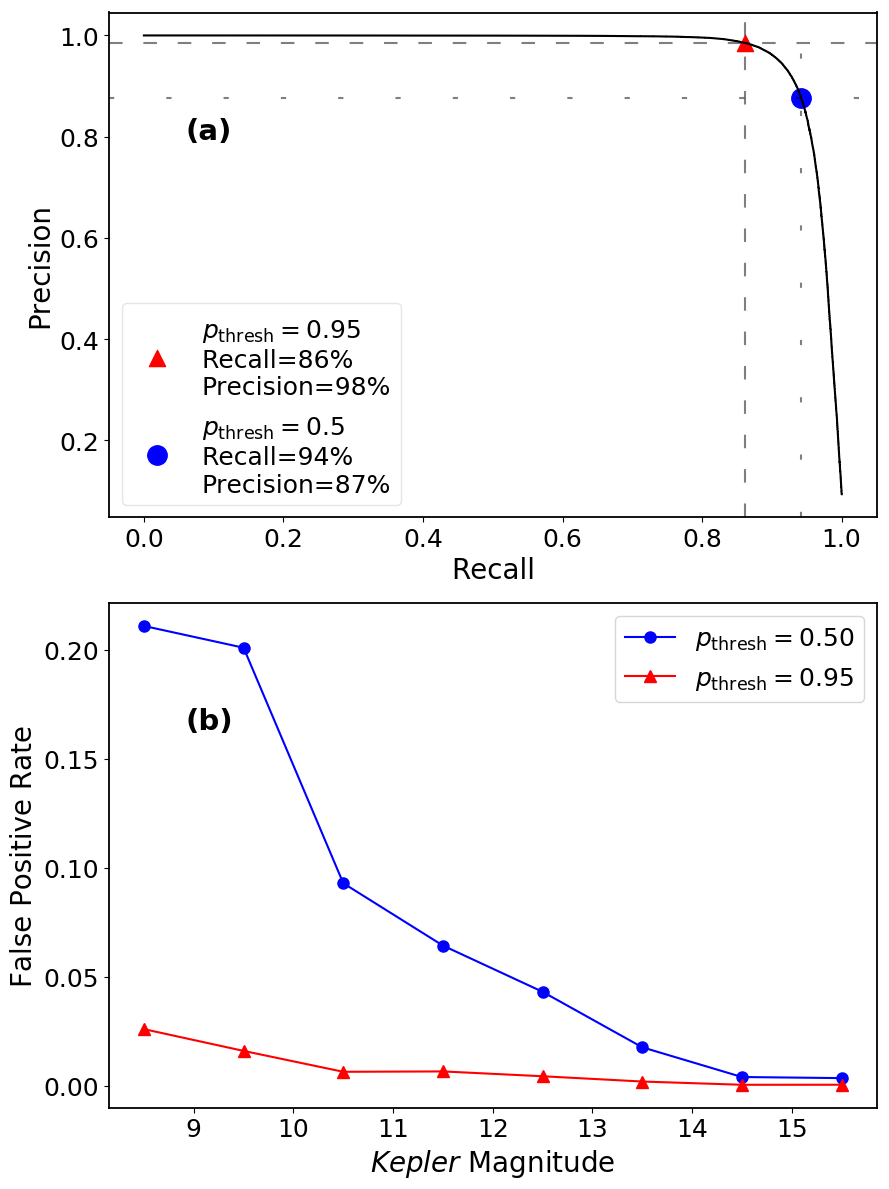}
    \caption{Performance of the classifier network on 27-day \textit{Kepler} test data. (a) Precision-recall plot for different acceptance thresholds of the network. Two particular thresholds are highlighted: a default threshold of $p_{\mathrm{thresh}}=0.5$ and a stricter threshold of $p_{\mathrm{thresh}}=0.95$ that we adopt in this study. (b) False positive rate as a function of \textit{Kepler} magnitude for different $p_{\mathrm{thresh}}$. The decrease towards fainter magnitudes suggests that most false positives in this test set are not from pure white noise spectra, but from spectra containing signals from other forms of variability like binarity or rotation, which can confuse the classifier.}
    \label{fig:val1}
\end{figure}

\subsubsection{Classification Thresholds}

Figure \ref{fig:val1}a demonstrates how changing the acceptance threshold for the classifier affects the recall (sample completeness) and precision (sample purity) of the resulting oscillating giant yield of our 27-day \textit{Kepler} test data. The recall metric reports the fraction of true oscillating giants within the test data that are successfully recovered by the classifier, while precision reports the fraction of accepted test stars that are truly oscillating giants. A lower acceptance threshold, $p_{\mathrm{thresh}}$, accepts more targets and thus has greater completeness but at the cost of lower purity (more false positives). Conversely, a higher $p_{\mathrm{thresh}}$ may miss a greater number of genuinely oscillating giants (lower recall) but is less likely to admit false positives. Indeed, Figure \ref{fig:val1}b shows that a strict threshold of $p_{\mathrm{thresh}}=0.95$ significantly reduces the false positive rate across all magnitudes in the test set. Because we expect raw QLP light curves to contain greater white noise levels and uncorrected instrumental noise compared to our \textit{Kepler} test set, we choose to be conservative in accepting positive detections and thus adopt $p_{\mathrm{thresh}}=0.95$.

\begin{figure}
    \centering
	\includegraphics[width=1.\columnwidth]{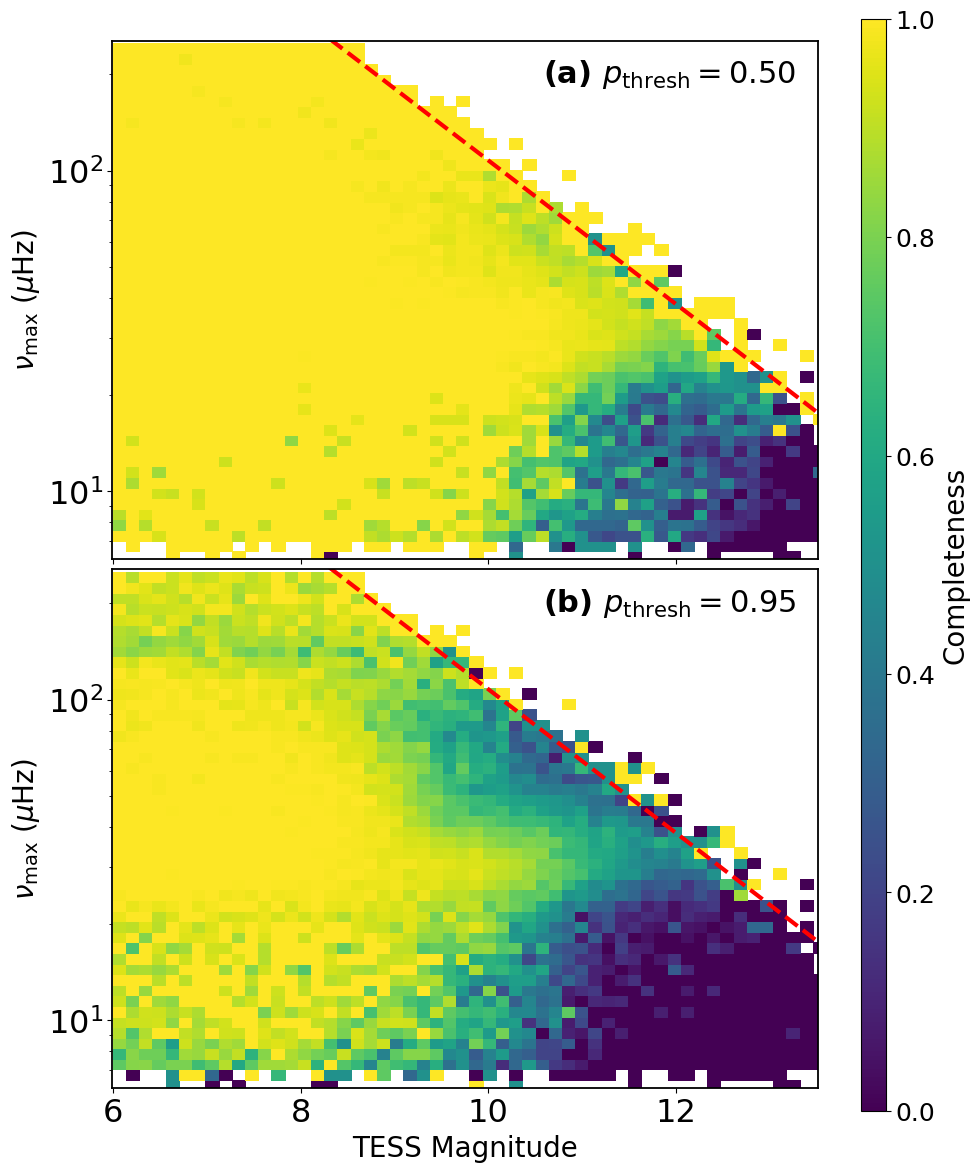}
    \caption{Completeness of the classifier on a test set of synthetic TESS red giants. (a) The recovery rate of oscillating red giants as a function of TESS magnitude and $\nu_{\mathrm{max}}$ when using an acceptance threshold, $p_{\mathrm{thresh}}=0.5$. (b) Similar to panel (a) but with $p_{\mathrm{thresh}}=0.95$. The red line in each panel delineates the boundary of the test set defined as where the height of the oscillation power excess is equal to ten times the white noise level in the power spectra.}
    \label{fig:val2}
\end{figure}

To assess how greater noise levels from TESS will affect our ability to detect oscillations, we evaluate the classifier's completeness on synthetic TESS-like light curves created using \texttt{celerite}\footnote{\href{https://celerite.readthedocs.io/}{https://celerite.readthedocs.io/}}\citep{celerite}, which models stochastic variability in astrophysical time series using Gaussian processes. 
Artificial red giant light curves are generated using current information available about oscillation mode parameters and the granulation background. The granulation background emulates `model F' from \citet{Kallinger_2014} with parameter values derived from the scaling relations found therein. The red giant universal pattern \citep{Mosser_2011} is used to generate frequencies of  $\ell=0,2,3$ oscillation modes from a given $\nu_{\mathrm{max}}$ value, with amplitudes estimated from the scaling relations given in \citet{Mosser_2012b} and the mixed $\ell=1$ oscillation modes computed according to the formalism by \citet{Vrard_2016} and \citet{Mosser_2018}. Because the mixed mode parameters depend on red giant evolutionary states, we determine evolutionary states for a synthetic red giant with a given $\nu_{\mathrm{max}}$ by sampling from a known $\nu_{\mathrm{max}}$-dependent distribution of known \textit{Kepler} red giant evolutionary states \citep{Elsworth_2019}. Next, we convert all derived oscillation parameters into the parameterization accepted by the simple harmonic oscillator kernel in \texttt{celerite} following \citet{Pereira_2019} and subsequently generate light curves of 27-day duration. To mimic the TESS observing window, we include a 1-day gap in the middle of the light curve, which simulates the data downlink event. Finally, we add white noise into the light curves following the TESS long-cadence noise properties described by \citet{Sullivan_2015}, where we assume observations with no instrumental noise contribution and an ecliptic latitude of 30$^{\circ}$.
Examples of simulated power spectra are shown in Appendix \ref{appendix:celerite_power_spectra}.


\begin{figure}
    \centering
	\includegraphics[width=1.\columnwidth]{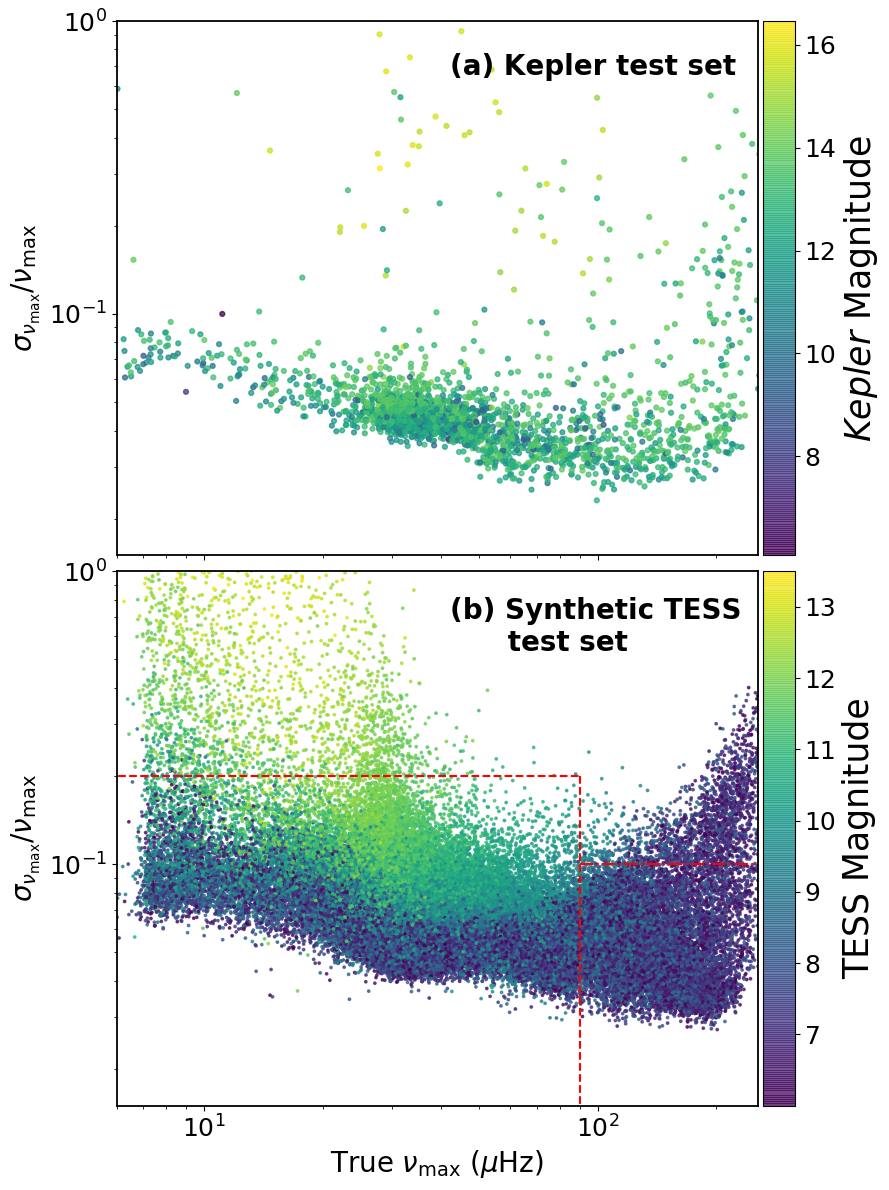}
    \caption{(a) The distribution of estimated fractional uncertainties of the $\nu_{\mathrm{max}}$ regression network for the \textit{Kepler} red giant test set, which comprises 15\% of the 16,194 red giants analyzed by \citet{Yu_2018} as described in Section \ref{sec:training_data}. (b) Same as panel (a) but for the simulated TESS test set. The red lines delineate boundaries that we use to indicate a confident estimate of $\nu_{\mathrm{max}}$.}
    \label{fig:val3}
\end{figure}

We simulate light curves across a uniform $\nu_{\mathrm{max}}$ range of [5, 250] $\mu$Hz and a uniform TESS magnitude range of [6.0, 13.5] at the same time. Figure \ref{fig:val2} shows the completeness of the classifier on a synthetic test set of 152,520 TESS red giants, where we can observe the performance of the network across a range of magnitudes and $\nu_{\mathrm{max}}$. As expected, using a larger $p_{\mathrm{thresh}}$ primarily affects the detection completeness of the test set near its upper boundary (red line), which is where the white noise levels in the simulated power spectra becomes comparable to the oscillation amplitudes. Both panels in Figure \ref{fig:val2} distinctly show low detection completeness for the faint and luminous giants, i.e., stars with TESS magnitudes fainter than 11 and $\nu_{\mathrm{max}}\apprle25\mu$Hz. As discussed previously by \citet{Hon_2018}, an important feature of a low $\nu_{\mathrm{max}}$ oscillating giant to the classifier is the visibility of the granulation profile, i.e., the slope of the red noise in power spectra. The high white noise levels for fainter stars significantly diminish the visibility of the granulation slope (see the examples in Appendix \ref{Appendix:RG_hard}), which makes the classifier less confident in identifying oscillations in such stars.

\subsubsection{Regression Threshold}
Additionally, we set a criterion based on the properties of the $\nu_{\mathrm{max}}$ regression network, which is trained independently of the classifier. Figure \ref{fig:val3} shows a comparison of fractional $\nu_{\mathrm{max}}$ uncertainties from the network across the \textit{Kepler} (real) and TESS-like (simulated) test sets in this study. The uncertainties shown for the \textit{Kepler} test set represent the best-case scenario for the network because the $\nu_{\mathrm{max}}$ regression network is trained on \textit{Kepler} data. Results on the synthetic test set, which has a broader range of white noise levels compared to its \textit{Kepler} counterpart, show a broader scatter of uncertainties \textemdash{} this includes a larger upwards scatter for fainter targets with $\nu_{\mathrm{max}}\apprle25\mu$Hz that we associate with the network's inability to locate a distinct oscillation power excess for such stars. We therefore determine a piecewise boundary (in red) below which we consider the network's $\nu_{\mathrm{max}}$ estimate to be consistent with a correct identification of an oscillation power excess. This boundary preserves over 96\% of the test set estimates in Figure \ref{fig:val3} and is described by $\sigma_{\nu_{\mathrm{max}}}/{\nu_{\mathrm{max}}}$ = 20\%  for $\nu_{\mathrm{max}}<90\mu$Hz and $\sigma_{\nu_{\mathrm{max}}}/{\nu_{\mathrm{max}}}$ = 10\% for $\nu_{\mathrm{max}}\geq90\mu$Hz.

\subsubsection{Threshold Summary} \label{section:threshold_summary}
The combined criteria that we use in our networks are as follows:

\begin{itemize}
    \item $p_{\mathrm{det}} \geq 0.95$ from the classifier.
    \item $\sigma_{\nu_{\mathrm{max}}}/{\nu_{\mathrm{max}}} \leq 20$\% for stars with $\nu_{\mathrm{max}}<90\mu$Hz.
    \item $\sigma_{\nu_{\mathrm{max}}}/{\nu_{\mathrm{max}}} \leq 10$\% for stars with $\nu_{\mathrm{max}}\geq90\mu$Hz.
\end{itemize}

\noindent These criteria are heuristic approaches that prioritize the \textit{precision} of the detection yield using our methods. We apply these concessions because we aim to report the highest quality detections from TESS QLP data to cater towards optimizing early spectroscopic follow-up across both northern and southern hemispheres of the sky.
Additionally, we assert that the recall and precision metrics reported in Figure \ref{fig:val1} are highly optimistic of our method's true performance on real TESS data. Their primary purpose is to guide us in the selection of $p_{\mathrm{det}}$ for our preliminary analysis over TESS QLP data. An important consequence of this is that our reported results in the forthcoming sections are not expected to represent a complete population of all oscillating giants from TESS FFIs. To properly assess the completeness of our networks in detecting oscillating giants, we require more principled approaches to calibrating decision thresholds by benchmarking our networks on real, labeled TESS data. Such a dataset has yet to be made because the systematic differences in photometric data quality between TESS and \textit{Kepler} have yet to be fully quantified, in addition to the recency of FFI light curves extracted in bulk. We will thus investigate the detection completeness of near all-sky TESS data in forthcoming work.

\begin{figure}
    \centering
	\includegraphics[width=1.\columnwidth]{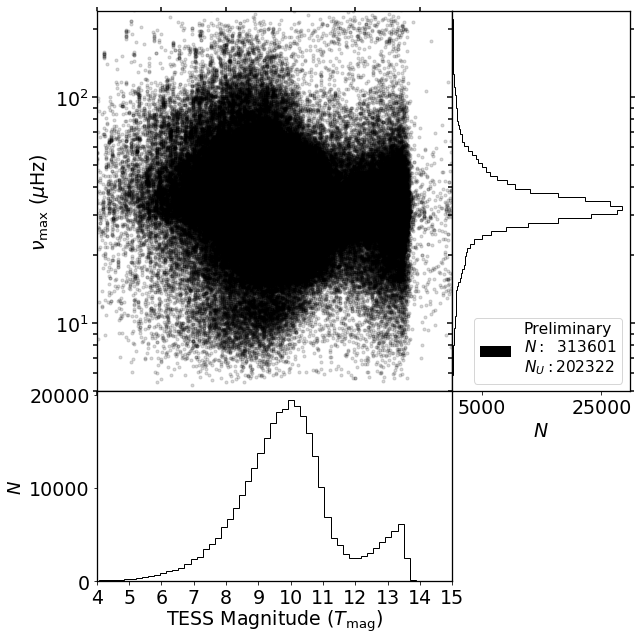}
    \caption{$\nu_{\mathrm{max}}$-magnitude distribution of all initial detections from the TESS QLP data. The number $N$ corresponds to the number of individual light curves whose power spectra show red giant oscillations, while the number $N_U$ corresponds to the total number of unique oscillating giants.}
    \label{fig:results1}
\end{figure}

\begin{figure}
    \centering
	\includegraphics[width=1\columnwidth]{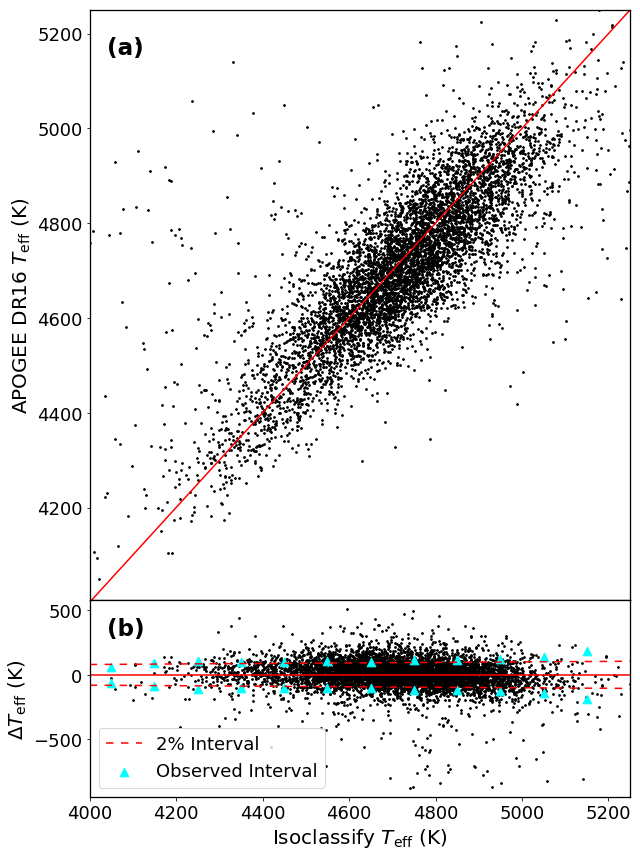}
    \caption{(a) A comparison between spectroscopic temperatures from APOGEE DR16 with the values measured in this work for 6,459 targets in our final yield in Section \ref{sec:final_sample}. The red line indicates the one-to-one relation. For visual clarity, we do not show outliers whose temperatures differences are beyond the 99.7th percentile range within this set of targets. (b) Residual differences in temperature defined as $\Delta T_{\mathrm{eff}}=$ (\texttt{isoclassify} $T_{\mathrm{eff}}$ $-$ APOGEE DR16 $T_{\mathrm{eff}}$). The dashed lines delineate a 2\% interval range representing the adopted \texttt{isoclassify} $T_{\mathrm{eff}}$ uncertainties. The cyan points delineate the observed uncertainty interval by calculating the dispersion of the residuals across different $T_{\mathrm{eff}}$ values.}
    \label{fig:temp}
\end{figure}

\section{Seismic Detection}
\label{sec:det}

\subsection{Preliminary Yield}
\label{section:low_luminosity_sample}
Figure \ref{fig:results1} shows the initial detection result from our neural networks comprising 202,322 unique FFI targets that meet the threshold criteria outlined in Section \ref{sec:threshold}. However, at this stage in our analysis, not all detected targets are guaranteed to be red giants because some are potentially faint non-red giants whose light curves show red giant oscillations due to flux contamination (`blending') from neighbouring giants. Due to the large TESS pixels, blending is expected to be particularly problematic for the detection task for fainter giants. Indeed, while the decline of detections at TESS magnitudes (T$_{\mathrm{mag}}$) fainter than 10 is expected especially for high $\nu_{\mathrm{max}}$ giants (c.f. Figure \ref{fig:training_data}), we observe an increase in the quantity of detected giants for T$_{\mathrm{mag}}\apprge11$ due to blending. We thus infer that the onset of significant blending in TESS occurs around T$_{\mathrm{mag}}\sim12$.




\subsection{Effective Temperatures and Radii}
\label{section:teff_rad}
To eliminate blended targets, we require that each target has a effective temperature ($T_{\mathrm{eff}}$) and radius ($R$) that is representative of an oscillating red giant. We calculate these stellar parameters for each target in our preliminary yield using the `direct method' of the \texttt{isoclassify}\footnote{\href{https://github.com/danxhuber/isoclassify}{https://github.com/danxhuber/isoclassify}} stellar classification code \citep{Huber_2017, Berger_2020}. The code uses parallaxes from \textit{Gaia} EDR3 \citep{Gaia_eDR3} as an input. These parallaxes are first corrected for zero-point offsets following the prescription by \citet{Lindegren_2020}, which applies corrections to stars with five- and six-parameter astrometric solutions. For stars brighter than a \textit{Gaia} magnitude of 10.8 that are corrected by the five-parameter offset model, we further subtract $0.015$mas from their corrected parallaxes following the recommendation by 
\citet{Zinn_2021}, who identified that the zero-point model over-corrects the \textit{Gaia} parallaxes for such stars through a comparison with \textit{Kepler} targets with known asteroseismic parallaxes. Using these corrected parallaxes, we then calculate distance posteriors using an exponentially decreasing volume density prior with a length scale of 1.35 kpc \citep{Bailer-Jones_2015, Astraatmadja_2016}.

We then calculate extinction values, $A_V$, for each star using the combined 3D dust maps of \citet{Drimmel_2003}, \citet{Marshall_2006}, and \citet{Green_2019} from the \texttt{mwdust} repository\footnote{\href{https://github.com/jobovy/mwdust}{https://github.com/jobovy/mwdust}}\citep{Bovy_2016}, where we assume an extinction uncertainty of 0.02 mag. These extinction values are applied to 2MASS $K$ band magnitudes to calculate apparent magnitudes. Bolometric corrections are inferred by interpolating $T_{\mathrm{eff}}$, $\log g$, [Fe/H], and  $A_V$ in the MESA Isochrones and Stellar Tracks \citep{Choi_2016} bolometric correction tables (MIST/C3K, Conroy et
al., in prep\footnote{\href{http://waps.cfa.harvard.edu/MIST/model\_grids.html}{http://waps.cfa.harvard.edu/MIST/model\_grids.html}}), where we
adopt an absolute solar bolometric magnitude of 4.74 mag, as appropriate to reproduce solar values for the bolometric corrections in the MIST grid. These bolometric corrections, which have an assumed uncertainty of 0.02 mag, are applied to the apparent magnitudes to derive absolute magnitudes and subsequently values of luminosities, $L$. Using 2MASS $JHK$ photometry, we apply the infrared flux method-based color-$T_{\mathrm{eff}}$ relations from \citet{GHB_2009} to infer values of $T_{\mathrm{eff}}$. Finally, we use $L$ and $T_{\mathrm{eff}}$ to estimate $R$ using the Stefan-Boltzmann relation.

A comparison between our temperature measurements with those from APOGEE Data Release 16 spectroscopy \citep[DR16]{Majewski_2017} in Figure \ref{fig:temp}a shows good levels of agreement and additionally demonstrates that the residual dispersion between both temperature sources are generally at the 2\% level (Figure \ref{fig:temp}b). We therefore approximate $\sigma_{\mathrm{T_{\mathrm{eff}}}}$ as 2\% of the reported $T_{\mathrm{eff}}$ value.
Meanwhile, output estimates for distances, $L$ and $R$ are reported as their distribution median values from Monte-Carlo sampling, with 1$\sigma$ confidence intervals derived from distribution percentiles. Fractional uncertainty estimates are typically 4\% for $L$ and 5\% for $R$.
With these fundamental stellar parameters, we now proceed with the analysis of our seismic sample. 

\begin{figure}
    \centering
	\includegraphics[width=1.\columnwidth]{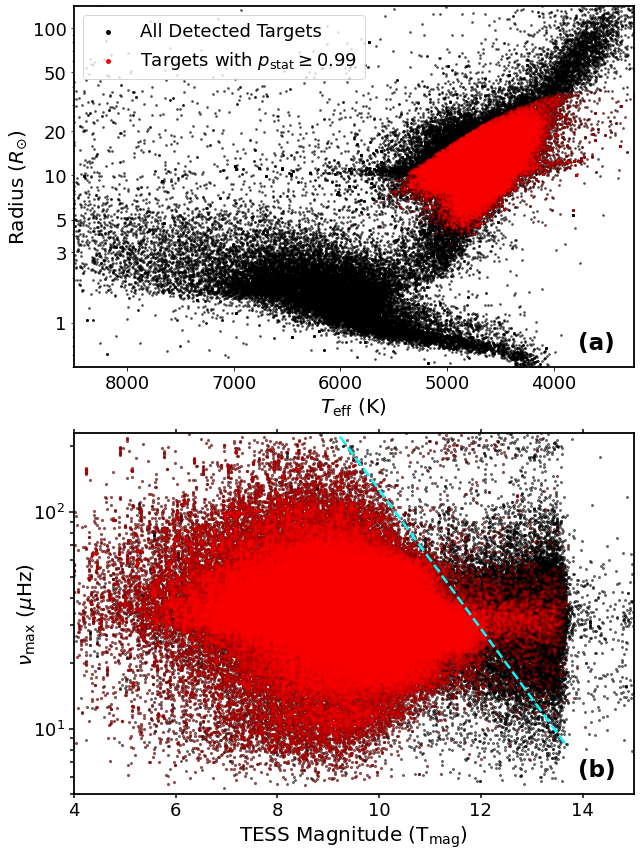}
    \caption{(a) Radius-temperature plot of all targets in the preliminary detection sample in Figure \ref{fig:results1} (black) and the remaining targets after retaining only those with $p_{\mathrm{stat}}\geq0.99$. (b) The  $\nu_{\mathrm{max}}$-magnitude distribution of the stars in panel (a). The blue line delineates a detection limit above which white noise levels are too high to detect visible red giant oscillations in power spectra. The deficiency of stars at $\nu_{\mathrm{max}}\apprle20\mu$Hz and T$_{\mathrm{mag}}\apprge11$ is caused by the limitations of our method as discussed in \ref{sec:threshold}.}
    \label{fig:results2}
\end{figure}

\begin{deluxetable*}{ccccccccc}[t]
\tablenum{1}
\tablecaption{List of 158,505 unique TESS targets comprising the final sample in this study. Included are each target's TESS magnitude (T$_\mathrm{mag}$), its estimated frequency at maximum oscillation power ($\nu_{\mathrm{max}}$), and measurements of effective temperature ($T_{\mathrm{eff}}$), radii ($R$), luminosity ($L$), and distances ($d$). Values of the re-normalized unit weight error (\texttt{ruwe}) from \textit{Gaia} EDR3 \citep{Gaia_eDR3} are included to flag potential non-single sources whose astrometric solutions may be problematic, which can affect the accuracy of our radius measurements. Values of \texttt{mass\_flag} are binary indicators of whether the target's stellar mass as estimated using the seismic scaling relation (Equation \ref{eqn:scaling_relation}) are within the `typical' mass range of a red giant from \textit{Kepler} (see text). The full version of this table is available in a machine-readable format in the online journal, with a portion shown here for guidance regarding its form and content. \label{table:low_luminosity}}
\tablewidth{0pt} 
\tablehead{
\colhead{TIC} & \colhead{$\nu_{\mathrm{max}}$} & \colhead{T$_\mathrm{mag}$} & \colhead{$T_\mathrm{eff}$} &
\colhead{$R$} & \colhead{$L$} & $d$ & \colhead{\texttt{ruwe}} & \colhead{\texttt{mass\_flag}}
\\
\colhead{} & \colhead{($\mu$Hz)} & \colhead{(mag)} & \colhead{(K)} &
\colhead{($R_{\odot}$)} & \colhead{($L_{\odot}$)} & \colhead{(kpc)} & \colhead{} & \colhead{}
}
\startdata
1078 & $30.8\pm2.2$&\hspace{1.4mm}9.7&$4629\pm92$&\hspace{-1.6mm}$11.3\pm0.5$& $52.6\pm2.3$&$0.828\pm0.012$&1.21 & 1\\
41959 & $14.9\pm2.2$&10.3&$4462\pm89$&\hspace{-1.6mm}$14.8\pm1.4$&\hspace{1.6mm}$77.5\pm13.0$&$1.333\pm0.109$&2.17 & 1\\
2026646 & $47.0\pm5.6$&12.8&$4415\pm88$&\hspace{-1.6mm}$16.3\pm1.2$&\hspace{1.6mm}$90.1\pm11.0$&$4.249\pm0.250$&0.95&0\\
31635521 & \hspace{-1.6mm}$112.1\pm6.1$ & \hspace{1.4mm}7.6 & $4909\pm98$ & $6.4\pm0.3$ & $21.5\pm0.7$ & $0.212\pm0.001$ & 1.10 & 1 \\
51931461 & $69.8\pm4.0$&\hspace{1.4mm}9.3&$4803\pm96$&$8.2\pm0.4$&$32.4\pm2.1$ &$0.554\pm0.015$&4.43&1\\
113288314 &\hspace{1.4mm}$7.8\pm1.0$&\hspace{1.4mm}9.7&$4092\pm81$&\hspace{-1.6mm}$24.6\pm1.1$&\hspace{-1.6mm}$152.1\pm7.2$ &$1.348\pm0.018$&0.97&1\\
150440218 & $87.5\pm5.4$ & 10.4 & $4815\pm96$ & $6.2\pm0.3$ & $18.8\pm0.7$ & $0.699\pm0.005$ & 0.97 & 1\\
$\cdots$&$\cdots$&$\cdots$&$\cdots$&$\cdots$&$\cdots$&$\cdots$&$\cdots$&$\cdots$\\
\enddata
\end{deluxetable*}

\subsection{Removing Blended Targets}
\label{section:remove_blends}
Having measured $T_{\mathrm{eff}}$ and $R$ for each target, we calculate their statistical detection probabilities, $p_{\mathrm{stat}}$, using the \citet{Chaplin_2011} global SNR estimation approach that has been adapted to TESS following the modifications described by \citet{Campante_2016} and \citet{Schofield_2019}.
More specifically, asteroseismic scaling relations are used to predict oscillation and granulation amplitudes \citep{Chaplin_2011, Kallinger_2014} from $T_{\mathrm{eff}}$ and $R$, while contributions from white noise and instrumental effects are estimated following the TESS instrumental specifications by \citet{Sullivan_2015}. These quantities are used to estimate the SNR of oscillations in a power spectrum and subsequently $p_{\mathrm{stat}}$. The value $p_{\mathrm{stat}}$ indicates the probability that the oscillation SNR exceeds a particular threshold, which \textemdash{} following \citet{Campante_2016}\textemdash{} is the SNR at which there exists a 5\% chance the detected oscillation signal arises from noise alone. Crucially, $p_{\mathrm{stat}}$ is a simple prediction of how likely oscillations can be observed for a given TESS target using only its fundamental stellar parameters and survey properties (e.g., apparent magnitude, observing cadence); it therefore does not directly use information from the observed power spectra of each target. As a result, our neural network detection algorithm is still needed to directly confirm the presence of visible oscillations from the observed data. Nonetheless, $p_{\mathrm{stat}}$ removes most of the obvious blends from our neural network detections. In particular, $p_{\mathrm{stat}}$ is small ($\ll$~0.99) for detected targets that either have $T_{\mathrm{eff}}$ and $R$ that are not representative of a red giant or TESS magnitudes too faint to expect detectable red giant oscillations.




When calculating $p_{\mathrm{stat}}$, we assume that each target experiences no blending (hence a flux dilution factor of 1) and a negligible systematic noise level ($\sigma_{\mathrm{sys}} = 0$). However, we do take into account the positional dependence of the background noise across the sky. To be conservative, we enforce a threshold of $p_{\mathrm{stat}}\geq0.99$, which as shown in Figure \ref{fig:results2}a, removes all blended targets and false positives within the instability strip (i.e., hot classical pulsators) and on the main sequence (dwarf stars). The $\nu_{\mathrm{max}}$-magnitude plot in Figure \ref{fig:results2}b shows that removing these false positives reveals a distinct boundary (blue line), which we parameterize as $\nu_{\mathrm{max}} = 10^{-0.32\cdot\mathrm{T}_{\mathrm{mag}}} + 5.3$. We interpret this boundary as the oscillation detection limit above which white noise levels are too high for oscillations to be detectable. In both Figures \ref{fig:results1} and \ref{fig:results2}b we note the lack of detected targets with $\nu_{\mathrm{max}}\apprle30\mu$Hz and T$_{\mathrm{mag}}\apprge11$ \textemdash{} we infer this observation to be caused by the limitation of our classifier as discussed in Section \ref{sec:threshold}, rather than a true deficiency of luminous giants in the underlying stellar population.
The remaining stars after the $p_{\mathrm{stat}}$ filtering comprises our final sample. Several examples of these stars documented in Appendix \ref{appendix:sample_examples}.

\section{Final Sample}
\label{sec:final_sample}
\subsection{Target Lists}
Table \ref{table:low_luminosity} lists 158,505 unique targets comprising the final sample together with their measured parameters in this study. Note that we report $\nu_{\mathrm{max}}$ measurements from only one sector for targets with multi-sector detections; we find this to be reasonable given that most multi-sector $\nu_{\mathrm{max}}$ measurements are consistent with the reported single-sector value. In particular, 97\% of $\nu_{\mathrm{max}}$ measurements across individual sectors of the same star deviate less than 2$\sigma$ of the quoted uncertainties in Table \ref{table:low_luminosity}.

We consider our yield of 158,505 oscillating giants to be a lower bound estimate of the yield of oscillating giants across TESS's nominal mission because of the conservative measures we have used in both detection and vetting methods. In comparison to the predicted all-sky yield of $\sim3\times 10^5$ as reported by \citet{Mackereth_2021}, our current yield suggests that there are still many giants that can potentially be detected from TESS. We defer more extensive detections across TESS FFIs to future work, which will better assess the sample completeness of red giants.

\begin{figure}
    \centering
	\includegraphics[width=1.\columnwidth]{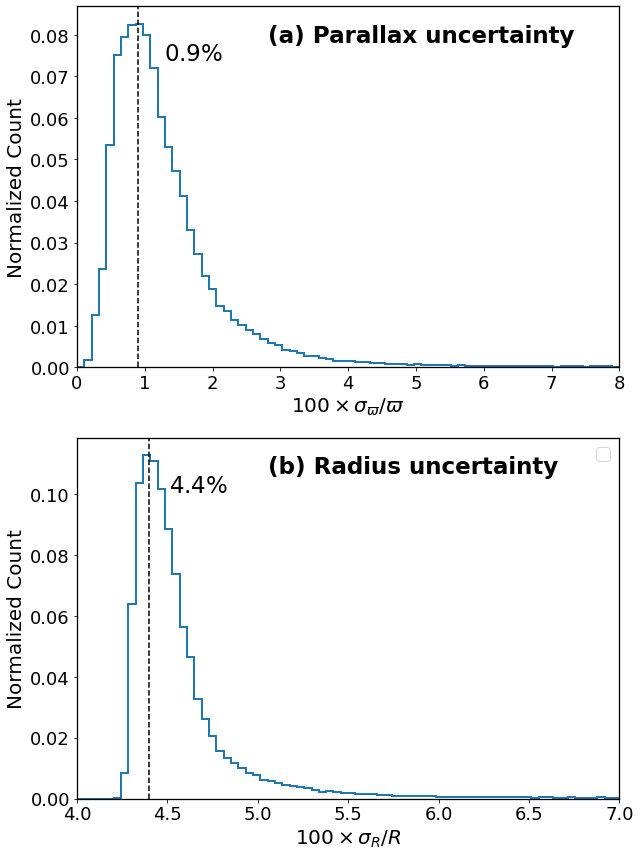}
    \caption{(a) The distribution of fractional parallax uncertainties ($\sigma_{\varpi}/\varpi$) for our final sample. The vertical line indicates a typical fractional uncertainty of 0.9\%. (b) The distribution of fractional radius uncertainties for our final sample, with the vertical line indicating typical radius uncertainties are about 4.4\%.}
    \label{fig:plx_rad}
\end{figure}

\begin{figure}
    \centering
	\includegraphics[width=1.\columnwidth]{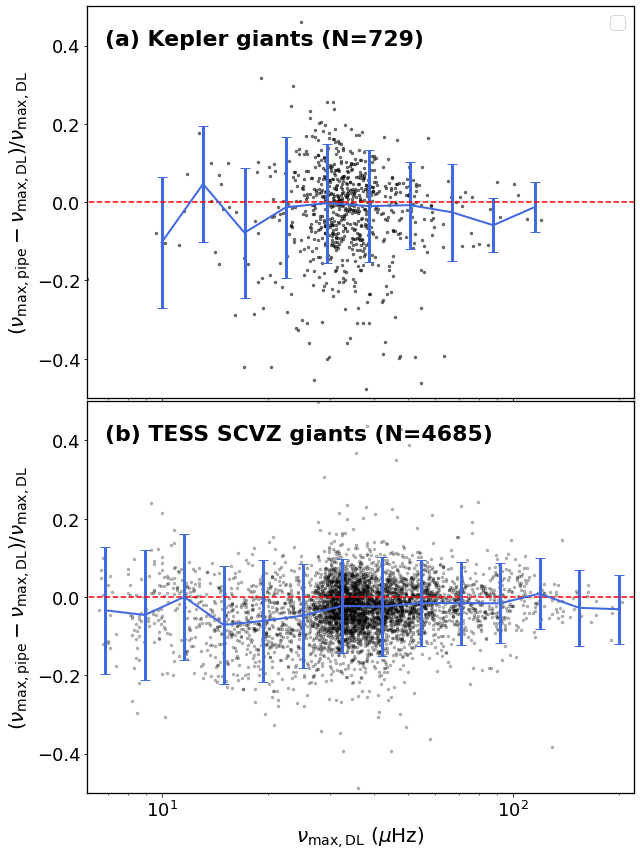}
    \caption{Comparisons of our estimates for the frequency at maximum oscillation power, ($\nu_{\mathrm{max, DL}}$), with measurements from asteroseismic data pipelines ($\nu_{\mathrm{max, pipe}}$) from independent studies of specific giant populations. The comparisons are for our detected targets that are: (a) in the \textit{Kepler} field \citep{Stello_2021}, and (b) in the Southern Continuous Viewing Zone \citep[TESS SCVZ]{Mackereth_2021}. The number $N$ indicates the total number of stars in each comparison, while the errorbars indicate the binned combined $\nu_{\mathrm{max}}$ uncertainties.}
    \label{fig:numax_comp}
\end{figure}

In Table \ref{table:low_luminosity}, we include values of the \textit{Gaia} re-normalized unit weight error \citep[\texttt{ruwe}]{Lindegren_2018}. Values of \texttt{ruwe}~$>1.40$ for a target indicate that the adopted astrometric solution is less likely to be reliable due to neighbouring companions (e.g., \citealt{Evans_2018, Belokurov_2020}), which can affect the accuracy of the target's reported value for $R$.

Although we have used measurements of $T_{\mathrm{eff}}$ and $R$ to filter out obvious blends in the final sample, such an approach cannot discern the occurrence of blending when both target and blending star are red giants. To potentially identify such scenarios, we additionally introduce a \texttt{mass\_flag} parameter that indicates whether or not the asteroseismic scaling mass ($M$) for each target falls within the `typical' mass range of a red giant. Following \citet{Stello_2008}, we estimate masses using the following: 
\begin{equation}
\label{eqn:scaling_relation}
    \frac{M}{M_{\odot}}=\left( \frac{\nu_{\mathrm{max}}}{\nu_{\mathrm{max, \odot}}} \right)\left(\frac{R}{R_{\odot}}\right)^2\left( \frac{T_{\mathrm{eff}}}{T_{\mathrm{eff, \odot}}} \right)^{0.5},
\end{equation}
with $\nu_{\mathrm{max, \odot}}=3090\mu$Hz \citep{Huber_2011} and $T_{\mathrm{eff, \odot}}=5772$K \citep{Prsa_2016}. Equation \ref{eqn:scaling_relation} is the seismic $\nu_{\mathrm{max}}$ scaling relation that relates the acoustic cutoff frequency of a star to its $M$, $R$, and $T_{\mathrm{eff}}$ \citep{Brown_1991, Kjeldsen_1995, Belkacem_2011}. 
We define the typical mass range of a red giant to be the values between the 0.5$^{\mathrm{th}}$ and 99$^{\mathrm{th}}$ percentile of $\sim$16,000 \textit{Kepler} red giant masses as measured by \citet{Yu_2018}, which corresponds to the interval 0.6$M_{\odot}$~$\geq M\geq$~2.9$M_{\odot}$. The vast majority of stars in our final sample are within this range and are assigned \texttt{mass\_flag}=1. There are a total of 3,415 ($\sim2.2\%$) stars in our sample with value outside the typical mass interval and are thus assigned \texttt{mass\_flag}=0 to indicate an outlier mass value.
We do not exclude these targets from our list because there are rare circumstances under which such extreme mass measurements can occur (such as a 3$M_{\odot}$ giant), but we caution that many of these are likely blends.

\subsection{Measurement Consistency}
In Figure \ref{fig:plx_rad}a, we show the uncertainty distribution for the parallaxes and radii estimates in our final sample. Given that our sample probes a relatively local region of the Galaxy (as further discussed in Section \ref{sec:results_ga}), the vast majority of our detected targets have precise astrometric measurements that result in a typical fractional radius uncertainty of 4.4\% as shown in Figure \ref{fig:plx_rad}b.
Additionally, we compare our $\nu_{\mathrm{max}}$ measurements from Tables \ref{table:low_luminosity}, $\nu_{\mathrm{max, DL}}$, with those from asteroseismic data pipelines, $\nu_{\mathrm{max, pipe}}$, from independent studies of TESS targets in specific sky regions. The comparison in Figure \ref{fig:numax_comp}a is for targets in the \textit{Kepler} field of view with observation lengths between 1-2 months based on the analysis by \citet{Stello_2021}. Meanwhile, the comparison in Figure \ref{fig:numax_comp}b is for targets in the TESS Southern Continuous Viewing Zone (TESS SCVZ) that comprise the `gold' sample from the \citet{Mackereth_2021} study \textemdash{} these typically have light curves with a duration of 3-12 months. In both comparisons, we find good levels of agreement between $\nu_{\mathrm{max, DL}}$ and $\nu_{\mathrm{max, pipe}}$, which suggests that our estimates of the frequency at maximum oscillation power are generally reliable.

\begin{figure*}[h]
    \centering
	\includegraphics[width=1.8\columnwidth]{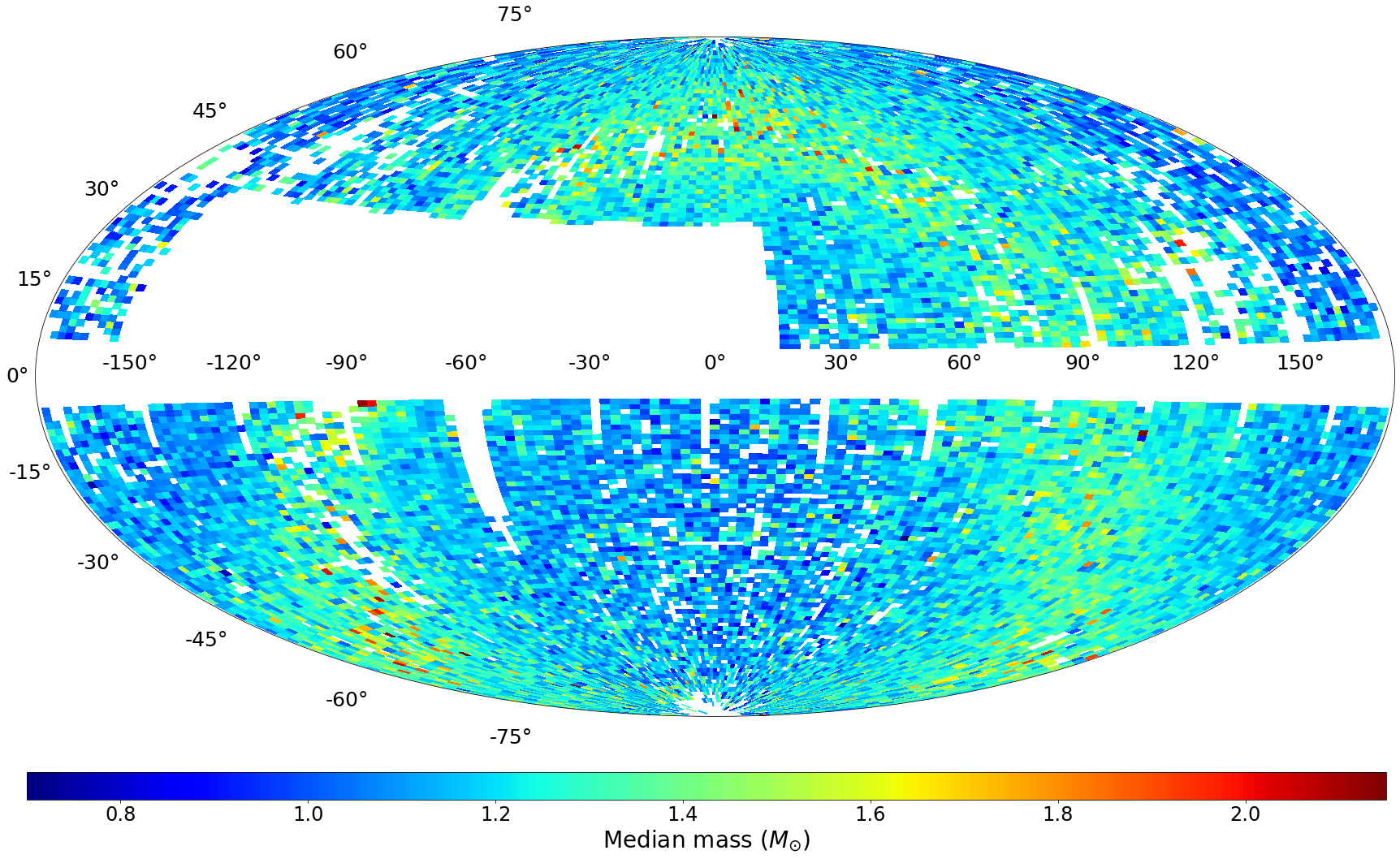}
    \caption{All-sky \textit{Gaia}-asteroseismology mass map of stars in the final sample with \texttt{ruwe} $\leq1.40$ and \texttt{mass\_flag}~$=1$ plotted in ecliptic coordinates. There are 150 bins across each dimension (longitude and latitude), with bins having fewer than 3 stars excluded. The missing patches of sky in the northern hemisphere correspond to Sectors 14-16 and 24-26, during which TESS's boresight was shifted towards higher ecliptic latitudes to avoid excessive stray Earth- and moonlight.}
    \label{fig:lowlum_GA_map1}
\end{figure*}
\begin{figure*}[h]
    \centering
	\includegraphics[width=1.8\columnwidth]{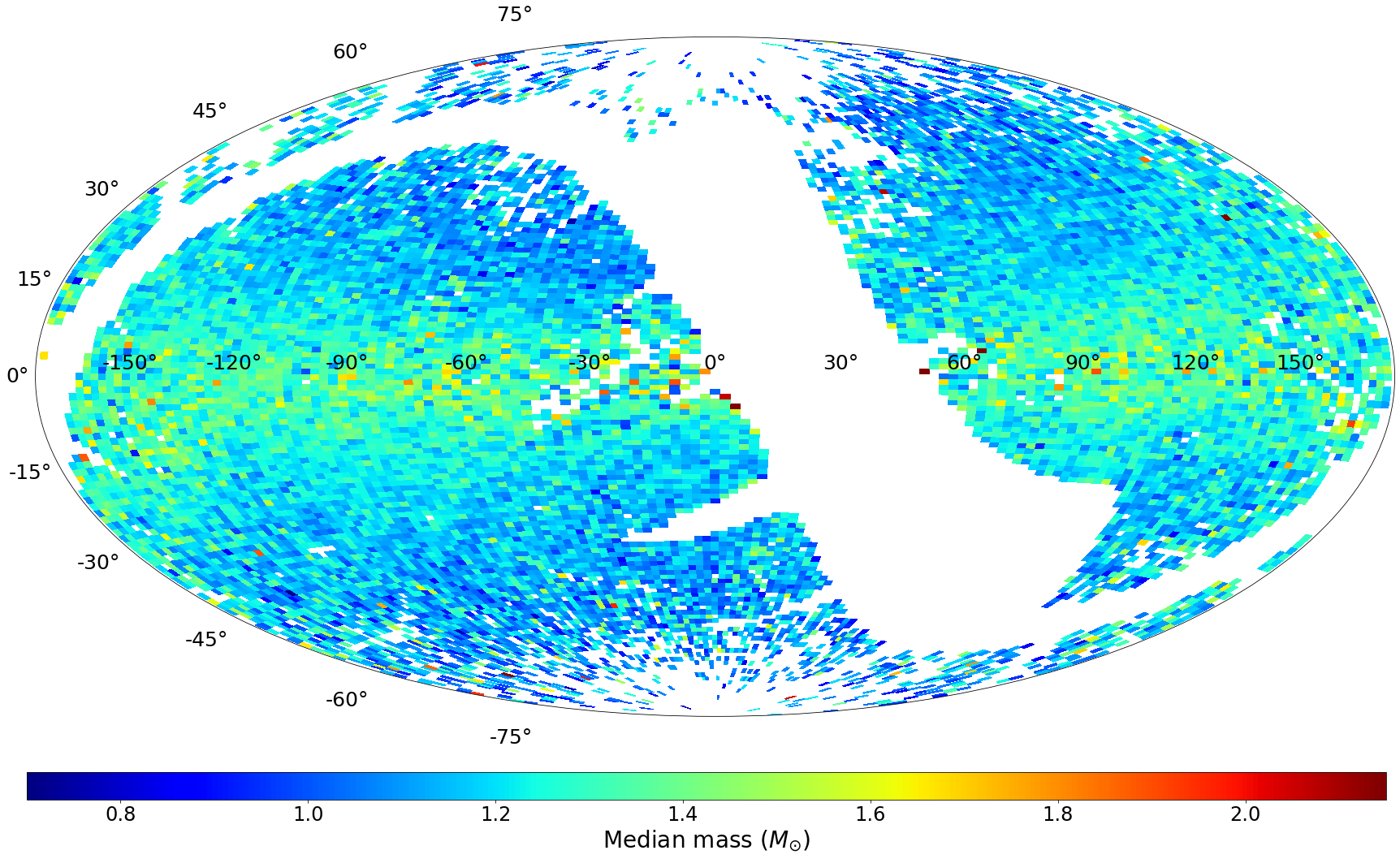}
    \caption{Same as Figure \ref{fig:lowlum_GA_map1}, but using the Galactic coordinate system ($l, b$).}
    \label{fig:lowlum_GA_map2}
\end{figure*}

\subsection{Caveats}
\label{sec:caveats}
There are two important additional caveats that should be considered with the current target lists:

\begin{enumerate}
    \item Blending may still be prevalent, especially when both the target and blending star have $\nu_{\mathrm{max}}$ values comparable to one another. We recommend more caution with the fainter targets (T$_{\mathrm{mag}}\sim12$) in our sample as well as those located closer to the Galactic plane, where crowding effects become more problematic. One approach that can inform the presence of blending, which we will explore in future work, is to systematically determine a contamination/crowding metric calculated across the photometric aperture of each target. Another possible method to identify giant-giant blending would be to determine if the stellar mass as estimated from Equation \ref{eqn:scaling_relation} (which uses \textit{Gaia}-derived radii) is consistent with a mass determined from the combined $\Delta\nu$-$\nu_{\mathrm{max}}$ scaling relation (e.g., \citealt[Equation 4]{Pinsonneault_2018}), where $\Delta\nu$ is the large frequency separation of the oscillation modes.
    \item The search performed in this study is not intended to be exhaustive, meaning that there are potentially many more oscillating giants from TESS that are yet to be detected. In the interest of producing an early, all-sky catalogue comprising strong candidates of oscillating giants, we make concessions to the population completeness of our final sample by vetting our preliminary yield with several thresholds. Because these thresholds are conservative, the number of stars in our final sample is expected to be a reasonable approximation of the minimum yield of oscillating giants that we can observe across the first two years of TESS.
\end{enumerate}


\section{Potential for Galactic Studies} 

\label{sec:results_ga}
\subsection{All-Sky Mass Map}
\label{sec:low_luminosity_sample_results}

In this section, we aim to demonstrate that our list of asteroseismic detections opens up prospects of inspecting broad spatial, chemical, and kinematic correlations over a near all-sky, continuous volume of the Milky Way using a sample of stars with their physical properties uniformly determined. At the same time, our demonstrations seek to provide an astrophysical validation of the accuracy of our results from a comparison with known physical trends from Galactic studies. To ensure that the measurements for our sample in the demonstrations are reliable,
we include only the 139,789 oscillating giants in our final sample with \texttt{ruwe}~$\leq$~1.40 and \texttt{mass\_flag}~=~1.


In Figures \ref{fig:lowlum_GA_map1} and \ref{fig:lowlum_GA_map2},
we show all-sky Hammer map projections of the final yield of our final sample, colored by seismic scaling masses derived using Equation \ref{eqn:scaling_relation}. Given that red giants follow an age-mass relation (e.g., \citealt{Miglio_2011, Bellinger_2020}), our results show that younger stars (higher mass stars) are mainly confined to the plane of the Galactic disc, i.e., the equator in Figure \ref{fig:lowlum_GA_map2}, with older stars (lower mass stars) generally populating higher Galactic latitudes. Notably, these trends are broadly consistent with global age trends obtained from large spectroscopic surveys \citep{Xiang_2017, Sanders_2018}. At $|l|\apprle120^{\circ}$ in Figure \ref{fig:lowlum_GA_map2}, we observe a slight trend of higher mass stars towards higher latitudes, which may indicate the presence of disk flaring (e.g., \citealt{Minchev_2015, Mackereth_2017}).

\subsection{Spatial Extent}

\begin{figure}
    \centering
	\includegraphics[width=1\columnwidth]{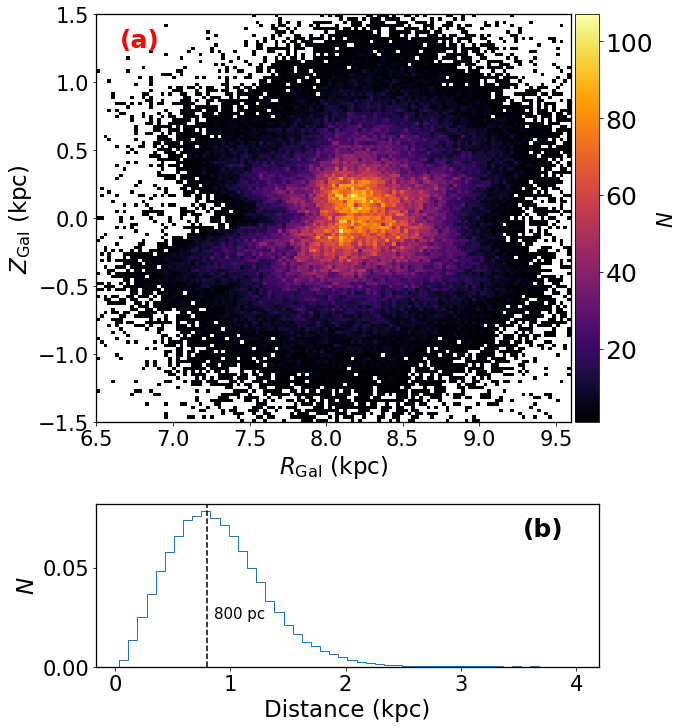}
    \caption{(a) Density map of oscillating giants detected in our study as a function of Galactocentric coordinates ($R_{\mathrm{Gal}}, Z_{\mathrm{Gal}}$). (b) Distance distribution of the stars in panel (a), with the mode of the distance distribution indicated by the line. }
    \label{fig:spatial_extent}
\end{figure}

We investigate the spatial distribution of our sample by calculating Galactocentric coordinates ($Z_{\mathrm{Gal}}, R_{\mathrm{Gal}}$) using the \texttt{Galactocentric} module from Astropy \citep{Astropy_2018}. We adopt a right-handed coordinate system that assumes a Galactic center position of $(X_{\mathrm{Gal}}, Y_{\mathrm{Gal}}, Z_{\mathrm{Gal}}) = (-8.122, 0, 0)$, such that the
Sun's radial distance from the Galactic center is 8.122~kpc and the Sun's height above the Galactic plane is taken to be 20.8~pc. Figure \ref{fig:spatial_extent} shows that our sample extends to distances primarily within 1~kpc indicating that our TESS yield is relatively local compared to \textit{Kepler} (e.g., \citealt{Rodrigues_2014, Mathur_2016}) or K2 (e.g., \citealt{Rendle_2019}). This outcome is expected given that the focus of the TESS mission is primarily on nearby, bright stars, and that our current yield is limited to TESS magnitudes brighter than 13.5. We additionally observe a lack of stars at $R_{\mathrm{Gal}}\apprle7.7$~kpc and $|Z_{\mathrm{Gal}}|\apprle0.1$~kpc. Stars within this region are positioned towards the highly crowded Galactic center, resulting in large levels of flux contamination that significantly lowers the detectability of red giant oscillations.

\begin{figure}
    \centering
	\includegraphics[width=1.\columnwidth]{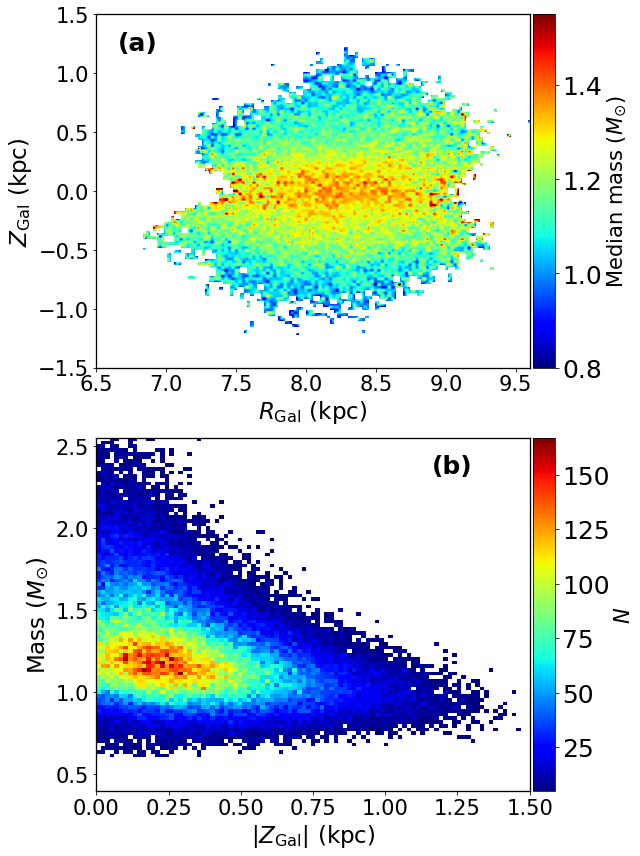}
    \caption{Mass distribution of our sample in Galactocentric coordinates. In each visualization, bins with fewer than 5 stars are not shown. (a) The map in Figure \ref{fig:spatial_extent}a colored by mass. (b) Distribution of masses as a function of vertical distance from the Galactic plane, $|Z_{\mathrm{Gal}}|$.}
    \label{fig:lowlum_coords}
\end{figure}

Figure \ref{fig:lowlum_coords}a shows the distribution of masses as a function of coordinates $(R_{\mathrm{Gal}}, Z_{\mathrm{Gal}})$. Similar to Figure \ref{fig:lowlum_GA_map2}, we find that more massive (and therefore young) stars near the Galactic plane at $Z_\mathrm{Gal}=0$. In this plot, a distinct dependence of masses with $Z_{\mathrm{Gal}}$ can be seen, which we further visualize in Figure \ref{fig:lowlum_coords}b. It is known that a vertical mass gradient of the Milky Way disk exists (e.g., \citealt{Casagrande_2016}), and therefore the large volume of our TESS yield presents the opportunity to probe the mass and age structure of the Galactic disk in greater detail.

\subsection{Kinematic Properties}

\begin{figure}
    \centering
	\includegraphics[width=1.\columnwidth]{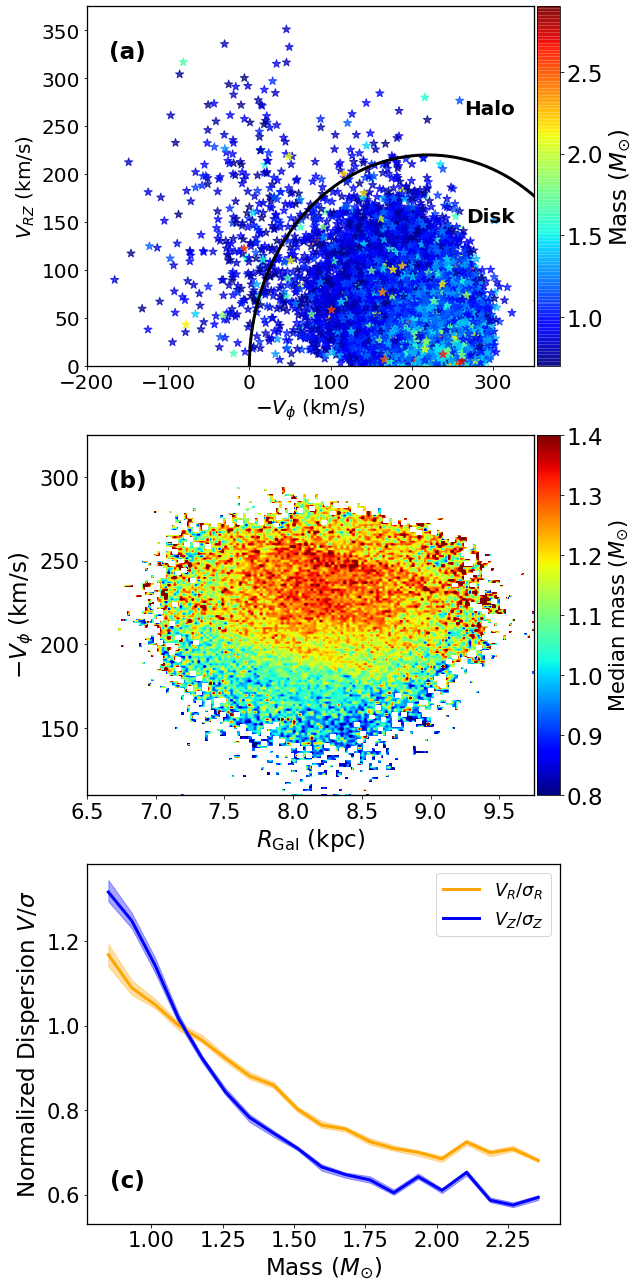}
    \caption{ Kinematic properties of the seismic sample in this study. (a) A Toomre diagram, with a line defined by $|V-V_{\mathrm{LSR}}|>220$~km/s corresponding to the boundary used by \citet{Bonaca_2017} to separate halo and disk populations, with $V_{\mathrm{LSR}}$ assumed as (0, 220, 0) km/s. (b) The phase space of our seismic sample, represented in the $(R_{\mathrm{Gal}}, V_{\phi})$ plane. In this plane, phase space substructures in the form of diagonal ridges that were first observed by \citet{Antoja_2018} can be seen. (c) Normlized radial and vertical velocity dispersions as a function of mass, computed following the \citet{Sharma_2020} approach. Uncertainties are indicated as the 16th and 84th percentile values in each mass bin, which are determined by bootstrapping.}
    \label{fig:toomre}
\end{figure}

To examine the kinematic properties of our red giant sample, 
we use parallaxes, radial velocities, and proper motions from \textit{Gaia} EDR3 to determine Galactocentric velocities $V=$~($V_{R}, V_{\phi}, V_Z$). Here, we consider a clockwise rotation of the Galaxy such that the azimuth angle $\phi=\tan(X_{\mathrm{Gal}}/Y_{\mathrm{Gal}})$ increases in the anti-clockwise direction.
Figure \ref{fig:toomre}a visualizes the distribution of our sample with a Toomre diagram, which is a useful representation for differentiating Galactic populations based on their kinematics (e.g., \citealt{Venn_2004}). We observe that most stars have $V_{\phi}\sim200-250$ km/s, which is approximately the circular velocity of the Local Standard of Rest (LSR, e.g., \citealt{Ding_2019}). In other words, most stars have velocities that are typical to objects orbiting the Milky Way within the disk. We additionally find a smaller population of stars with atypical disk kinematics \textemdash{} these stars may have larger combined radial and vertical velocities ($V_{RZ}$) compared to those in the disk as well as highly retrograde motion relative to the LSR. Such stars are identified as having kinematics representative of a Galactic halo population; for instance \citet{Bonaca_2017} identifies stars with $|V-V_{\mathrm{LSR}}>220|$~km/s (outside the line in Figure \ref{fig:toomre}) as belonging to the halo. Using this criterion, we identify 354 potential halo candidates in our sample, which carries forward prospects of probing the Milky Way halo with TESS in subsequent work. 

Figure \ref{fig:toomre}b shows the kinematic distribution of our sample in the $(R_{\mathrm{Gal}}, V_{\phi})$ plane, where we can observe the phase-space substructures in the form of diagonal ridges that were discovered by \citet{Antoja_2018}. The study by \citet{Khanna_2019} demonstrated that the stars forming such substructures are primarily those
that are close to the Galactic plane (small $|Z_{\mathrm{Gal}}|$) and suggested that they are either due to: a) kinematic perturbations from the phase-mixing of transient spiral arms; b) the interactions between low $|Z_{\mathrm{Gal}}|$ stars that may have been externally perturbed by a satellite such as a dwarf galaxy; c) stars near the Galactic plane being kinematically `cold' and thus easier to perturb. In Figure \ref{fig:toomre}b, we indeed find that typically stars with higher masses (younger) show the diagonal ridges more prominently, which is consistent with the results by
\citet{Khanna_2019} that these stars belong to regions near the Milky Way plane.

We provide another astrophysical validation of our seismic sample using velocity dispersions in Figure \ref{fig:toomre}c. Using the \citet{Sharma_2020} approach, we calculate radial ($\sigma_R$) and vertical ($\sigma_Z$) velocity dispersion profiles solely as a function of mass under a solar metallicity approximation. The observed decrease of the dispersion profiles with mass is consistent with the expectation that velocity dispersions are expected to increase with age due to dynamical heating over time from sources such as giant molecular clouds (e.g., \citealt{Lacey_1984}) and spiral arms (e.g., \citealt{Barbanis_1967}). Qualitatively, the greater steepness of the $(V_Z/\sigma_Z)$ profile compared to that of $(V_R/\sigma_R)$ in Figure \ref{fig:toomre}c agrees with the observational results by \citet{Sharma_2020} that suggests a high efficiency of heating by giant molecular clouds (a significant contributor to vertical scattering) early in the dynamical history of the Galaxy \textemdash{} this is notably a result predicted by the simulations by \citet{Aumer_2016}. 

Based on theory, the asymmetric drift \citep{Stromberg_1946}, namely how much a star's circular motion ($-V_{\phi}$) lags behind the $V_{\mathrm{LSR}}$, is larger with for stars with high $\sigma_R$ because such kinematically hot stars require less centrifugal support to remain in dynamical equilibrium (e.g., \citealt{Binney_2008, Sharma_2014}). Therefore, the greater the difference between $-V_{\phi}$ and $V_{\mathrm{LSR}}=(0, 220, 0)$ km/s in Figures \ref{fig:toomre}a and \ref{fig:toomre}b, the larger the asymmetric drift. Because $\sigma_R$ scales with age (and mass, as per Figure \ref{fig:toomre}c), we should expect a reduction in the average stellar mass as $-V_{\phi}$ decreases below 220 km/s. Indeed, such a trend is observed in Figure \ref{fig:toomre}b, which again shows the consistency of the masses of our sample from a kinematic perspective.  



\subsection{Chemical Properties}

\begin{figure}
    \centering
	\includegraphics[width=1.\columnwidth]{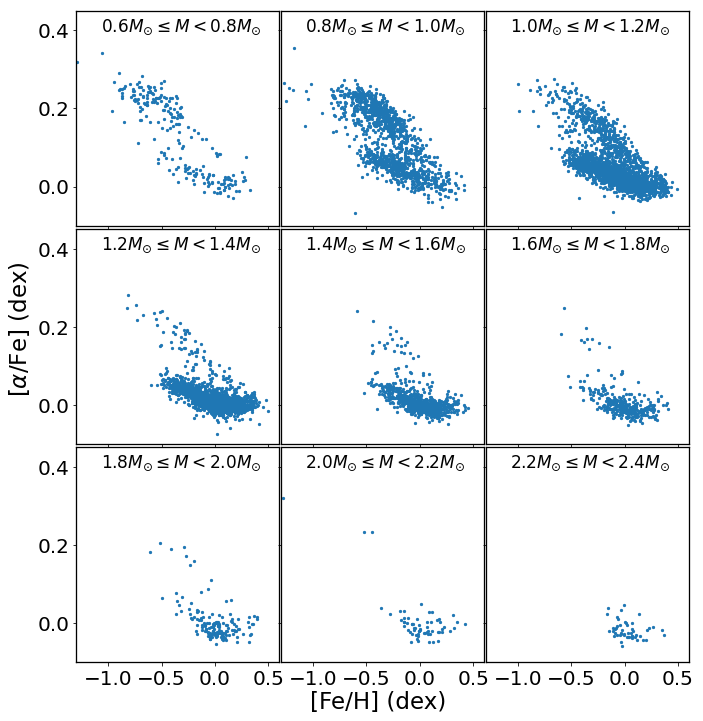}
    \caption{The [$\alpha$/Fe]-[Fe/H] diagram of 6,459 stars in our sample that have chemical abundances from APOGEE Data Release 16. Different panels correspond to different mass bins.}
    \label{fig:chem}
\end{figure}

Figure \ref{fig:chem} shows the [$\alpha$/Fe]-[Fe/H] diagram for 6,459 stars with measured chemical abundances from APOGEE Data Release 16 \citep{Majewski_2017} of the Sloan Sky Digital Survey \citep{Ahumada_2020}. The [$\alpha$/Fe]-[Fe/H] relation is a frequently examined diagnostic in chemical evolution studies of the Milky Way disk (e.g., \citealt{Adibekyan_2011,Haywood_2013, Bensby_2014, Hayden_2015, Anders_2017, Aguirre_2018}), which reveals two distinct stellar populations from a bi-modality in the observed [$\alpha$/Fe] abundances in Figure \ref{fig:chem}. The first is a population of solar-[$\alpha/$Fe] stars spanning a wide range of metallicities \textemdash{} these are expected to comprise young stars confined to the plane of the Galaxy  (e.g., \citealt{Hayden_2015}). Assuming mass as an age proxy, we observe that most stars with $M\ge1.6M_{\odot}$ (thus typically young) in Figure \ref{fig:chem} have solar-[$\alpha/$Fe] abundances, in agreement with expectations. However, about $\sim$20 targets within these high-mass bins have [$\alpha/$Fe] $\sim0.3$ dex \textemdash{} such stars are potentially the peculiar `young-$\alpha$ rich' stars \citep{Chiappini_2015, Martig_2015} whose features are not predicted by standard chemical evolution models of the Galaxy.

The second distinct population in Figure \ref{fig:chem} is seen at [$\alpha$/Fe]~$\apprge0.15$~dex, which corresponds to the high-$\alpha$ sequence. This populations is typically associated with metal-poor stars with larger vertical scale heights compared to those in the low-$\alpha$ population (e.g., \citealt{Bovy_2012}). Although stars in the high-$\alpha$ sequence were previously found to strongly correlate with age \citep{Haywood_2013}, more recent studies present evidence to the contrary from the identification of old, metal-rich low-$\alpha$ stars (e.g., \citealt{Bensby_2014, Aguirre_2018}).  We find stars with $M\apprle1.0M_{\odot}$ (typically older) in Figure \ref{fig:chem} to be distributed in both low- and high-$\alpha$ populations, which does support the presence of old, metal-rich stars. While a detailed investigation in these chemical trends is beyond the scope of this paper, we aim to show that our seismic sample provides great utility not only in revealing trends in chemical abundances, but also in potentially identifying a subset of chemically interesting targets within large stellar populations.

\section{Conclusions}
We have presented the first all-sky sample of oscillating red giants using MIT Quick-Look Pipeline photometry of the first 26 sectors of TESS Full Frame Images. Our main results are summarized as follows:

\begin{itemize}
    \item We reported red giant oscillations in a total of 158,505 targets from over 20 million light curves spanning TESS magnitudes as faint as 13.5 using neural networks. The detected giants have frequencies at maximum oscillation power ($\nu_{\mathrm{max}}$) spanning 6$\mu$Hz$~\apprle\nu_{\mathrm{max}}\apprle250\mu$Hz, which corresponds to a surface gravity range of 1.5 dex $\apprle\log(g)\apprle3.0$ dex. Due to the limitations of the neural network classifier, our reported yield does not currently include the most faint and luminous targets, which are those with TESS magnitudes $\apprge11$ and $\nu_{\mathrm{max}}\apprle15\mu$Hz. 
    
    \item For each seismic detection, we reported estimates of its frequency at maximum power ($\nu_{\mathrm{max}}$), luminosity ($L$), radius ($R$), effective temperature ($T_{\mathrm{eff}}$), and distance ($d$). To produce a high quality target list of oscillating red giants, we have applied thresholds in our machine learning algorithms and applied statistical detection probabilities based on global SNR estimates to minimize the number of false positives. Our reported results are therefore that of an early detection yield and not that of a complete survey of all underlying oscillating giants from TESS Full Frame Images. Due to (1) our use of conservative vetting measures, (2) the lack of optimal light curve corrections required for asteroseismology, and (3) the use of only one month-long observations, it is reasonable to expect that the number of targets in our reported detection yield is a lower bound to the complete set of oscillating red giants that can be obtained from TESS's nominal mission.
    
    \item With our detection yield we demonstrated the enormous potential of TESS for all-sky Galactic archeology by presenting the first near all-sky \textit{Gaia}-asteroseismology mass map. With data from the Quick-Look Pipeline limited to a TESS magnitude of 13.5, we found our yield to typically span distances within 1~kpc. Despite being limited to a relatively local volume of the Galaxy, our yield provides the most comprehensive distribution of seismic measurements across the Galaxy to date. Our measurements, when applied to spatial, kinematic, and chemical studies of the Galaxy, show good qualitative agreement with expected astrophysical trends as a function of mass. Specifically, we showed that 1) the masses of stars within the Galactic plane is on average larger than those farther from the plane; 2) a vertical mass gradient of the Milky Way disk can be distinctly observed within our sample; 3) there are 354 oscillating giants in our sample that are have Galactic halo-like kinematics; 4) stars prominently showing kinematic phase-space ridges are on average those that are higher in mass and reside within the Galactic plane; 5) lower mass stars typically have larger radial and vertical velocity dispersions, consistent with the interpretation that such stars are older; 6) there exists distinctions between low-$\alpha$ and high-$\alpha$ populations as a function of stellar mass in our sample.
\end{itemize}

With a good qualitative agreement with trends from Galactic studies using masses as a proxy for stellar age, it is reasonable to expect that the use of actual age estimates from a combination of TESS all-sky asteroseismology, \textit{Gaia} parallaxes, and large survey spectroscopy will provide very powerful probes of Milky Way evolution.
Finally, we note that our `quick look' at the all-sky asteroseismic yield from TESS has provided insights on how to improve our data and methodology for further work to expand the yield across Full Frame Images. Currently, we have used only `raw' data from the Quick-Look Pipeline, which lack detrending to remove instrumental noise. Upcoming data releases from the TESS Asteroseismic Science Consortium (e.g., \citealt{TASOC_Data_Release_5}) will provide light curves with instrumental noise corrections that are optimized for asteroseismology for all FFI targets up to a TESS magnitude of 15, which will significantly increase the detection yield across all TESS sectors. Another vital insight is the need for improved machine learning algorithms and training data. As the asteroseismic analyses of TESS data further progresses across the scientific community, it will become possible to use or develop training data that is more representative of real TESS FFI observations. The availability of better training data will subsequently facilitate improvements with our deep learning algorithms, particularly in the detection of oscillations for the faintest and most luminous targets from TESS. These future prospects, when combined with TESS's potential for all-sky Galactic studies, will certainly provide a more complete and detailed view of the Milky Way.



\begin{acknowledgments}
Funding for the TESS mission is provided by the NASA Explorer Program. 
M.H. acknowledges support from NASA through the NASA Hubble
Fellowship grant HST-HF2-51459.001 awarded by the Space Telescope
Science Institute, which is operated by the Association of Universities for Research in Astronomy, Incorporated, under NASA contract NAS5-26555.
D.H. acknowledges support from the Alfred P. Sloan Foundation, the National Aeronautics and Space Administration (80NSSC21K0652), and the National Science Foundation (AST-1717000). JT acknowledges support for this work was provided by NASA through the NASA Hubble Fellowship grant \#51424 awarded by the Space Telescope Science Institute, which is operated by the Association of Universities for Research in Astronomy, Inc., for NASA, under contract NAS5-26555 and by NASA Award 80NSSC20K0056. JCZ is supported by an NSF Astronomy and Astrophysics Postdoctoral Fellowship under award AST-2001869. MV acknowledges support from NASA grant 80NSSC18K1582.
\end{acknowledgments}

\vspace{5mm}


\software{Astropy \citep{Astropy_2018}, \texttt{celerite} \citep{celerite}, \texttt{isoclassify} \citep{Huber_2017, Berger_2020}, Pytorch \citep{Pytorch} 
          }



\appendix

\section{Network Architecture} \label{Appendix:Network_Architecture}
Tables \ref{table:classifier_structure} and \ref{table:regressor_structure} detail the structure of the classifier and the regression network, respectively. The networks are developed using the Pytorch version 1.1.0 deep learning library \citep{Pytorch}. The networks are trained using the Adam optimizer \citep{Kingma_2014} with an initial learning rate of 0.0001 and early stopping applied.

The networks are generally similar to those implemented by \citet{Hon_2018}, with a few key differences. First, both networks have dropout applied only to the final feature extraction layer, with a small dropout probability of 0.1. Because this current work has significantly more training data compared to the \textit{Kepler}-as-K2 used by \citet{Hon_2018}, weaker regularization is required to prevent network overfitting. The second difference relates to the uncertainty estimation of the $\nu_{\mathrm{max}}$ regression network. The previous study uses Monte Carlo dropout \citep{Gal_2016}, while our work in this study explicitly estimates $\sigma_{\mathrm{\nu_{\mathrm{max}}}}$ through an auxiliary output layer (e.g., \citealt{Kendall_2017}). Consequently, optimizing the $\nu_{\mathrm{max}}$ regression network is not done by minimizing the mean squared error (as before), but the negative log-likelihood, $E$, given by the following:

\begin{equation}\label{eq:Gaussian_likelihood}
    p(y\mid \mathbf{x}) =  \frac{1}{(2\pi)^{1/2}\sigma(\mathbf{x})}\exp{\left(- \frac{(y-\nu_{\mathrm{max, pred}})^2}{2\sigma(\mathbf{x})^2}\right)},
\end{equation}

\begin{equation}\label{eq:nll}
    E = \sum_{m=1}^{m_\mathrm{tot}}-\ln{p(y_m \mid \mathbf{x}_m)}.
\end{equation}  

Here, $\mathbf{x}$ is the input image, $\sigma(\mathbf{x})$ is $\sigma_{\mathrm{\nu_{\mathrm{max}}}}$, $y$ is the ground truth value of $\nu_{\mathrm{max}}$, $\nu_{\mathrm{max, pred}}$ is the estimated value of the frequency at maximum power by the $\nu_{\mathrm{max}}$ regression network, and $m_\mathrm{tot}$ is total number of power spectra in the training set.  

\begin{deluxetable}{cccc}
\tablenum{1A}
\tablecaption{Structure of the classifier to detect oscillating red giants from images of power spectra. \label{table:classifier_structure}}
\tablewidth{0pt} 
\tablehead{
\colhead{Component} & \colhead{Layer} & \colhead{Filter Size} & \colhead{Output Shape}
} 
\startdata
{       }& \texttt{conv1}$^b$ & (5,5) & (128,128,4) \vspace{0.075cm}\\
{       } & \texttt{pool1} & (2,2) & (64,64,4) \vspace{0.075cm}\\
{       } & \texttt{conv2} & (3,3) & (64,64,8) \vspace{0.075cm}\\
Feature Extraction & \texttt{pool2} & (2,2) & (32,32,8) \vspace{0.075cm}\\
{       } & \texttt{conv3} & (3,3) & (32,32,16) \vspace{0.075cm}\\
 {       }& \texttt{pool3} & (2,2) & (16,16,16) \vspace{0.075cm}\\
 {       }& \texttt{flatten} & - & (4096,)\vspace{0.075cm}\\
{       } & \texttt{drop}$^c$ & - & (4096,) \vspace{0.075cm}\\
 \hline 
\multirow{2}{2cm}{\centering Class Score Estimation}  & \texttt{dense1} & (4096,128) & (128,) \vspace{0.075cm}\\ 
 & \texttt{output} & (128,1) & (1,) \vspace{0.075cm}\\
\enddata
\tablecomments{The rectified linear unit activation \citep[ReLU]{Nair_2010} is applied after every \texttt{conv} and \texttt{dense} layer. \newline$^a$ For convolutional layers, this column corresponds to the dimensions of the convolving kernel, whereas for dense layers the column corresponds to (\texttt{number of neurons in previous layer}, \texttt{number of neurons in current layer}). \newline $^b$ For convolutional layers, weight shapes are in format (number of filters, receptive field size), while output shapes are in format (height, width, number of filters). \newline $^c$ We implement regular Dropout \citep{Hinton_2012, Sriva_2014} with a drop probability of 0.1.}
\end{deluxetable}

\begin{deluxetable}{cccc}
\tablenum{1B}
\tablecaption{Structure of the regression network to estimate $\nu_{\mathrm{max}}$ from images of power spectra showing red giant oscillations. \label{table:regressor_structure}}
\tablewidth{0pt} 
\tablehead{
\colhead{Component} & \colhead{Layer} & \colhead{Filter Size} & \colhead{Output Shape}
} 
\startdata
{       }& \texttt{conv1}$^b$ & (5,5) & (128,128,4) \vspace{0.075cm}\\
{       } & \texttt{pool1} & (2,2) & (64,64,4) \vspace{0.075cm}\\
{       } & \texttt{conv2} & (3,3) & (64,64,8) \vspace{0.075cm}\\
Feature Extraction & \texttt{pool2} & (2,2) & (32,32,8) \vspace{0.075cm}\\
{       } & \texttt{conv3} & (3,3) & (32,32,16) \vspace{0.075cm}\\
 {       }& \texttt{pool3} & (2,2) & (16,16,16) \vspace{0.075cm}\\
 {       }& \texttt{flatten} & - & (4096,)\vspace{0.075cm}\\
{       } & \texttt{drop}$^c$ & - & (4096,) \vspace{0.075cm}\\
 \hline 
 \multirow{2}{2cm}{\centering Mean Value Estimation} & \texttt{dense1a} & (4096,256) & (256,) \vspace{0.075cm}\\ 
& \texttt{dense2a} & (256,256) & (256,) \vspace{0.075cm}\\
& \texttt{output} $\nu_{\mathrm{max}}$ & (256,1) & (1,) \vspace{0.075cm}\\
\hline \vspace{0.075cm}
\multirow{2}{2cm}{\centering Uncertainty Estimation} & \texttt{dense1b} & (4096,256) & (256,) \vspace{0.075cm}\\ 
& \texttt{dense2b} & (256,256) & (256,) \vspace{0.075cm}\\
& \texttt{output} $\sigma_{\nu_{\mathrm{max}}}$ & (256,1) & (1,) \vspace{0.075cm}\\
\enddata
\tablecomments{The rectified linear unit activation \citep[ReLU]{Nair_2010} is applied after every \texttt{conv} and \texttt{dense} layer. \newline$^a$ For convolutional layers, this column corresponds to the dimensions of the convolving kernel, whereas for dense layers the column corresponds to (\texttt{number of neurons in previous layer}, \texttt{number of neurons in current layer}). \newline $^b$ For convolutional layers, weight shapes are in format (number of filters, receptive field size), while output shapes are in format (height, width, number of filters). \newline $^c$ We implement regular Dropout \citep{Hinton_2012, Sriva_2014} with a drop probability of 0.1.}
\end{deluxetable}

\section{\texttt{celerite} Power Spectra}
\label{appendix:celerite_power_spectra}
\begin{figure}
    \centering
	\includegraphics[width=\columnwidth]{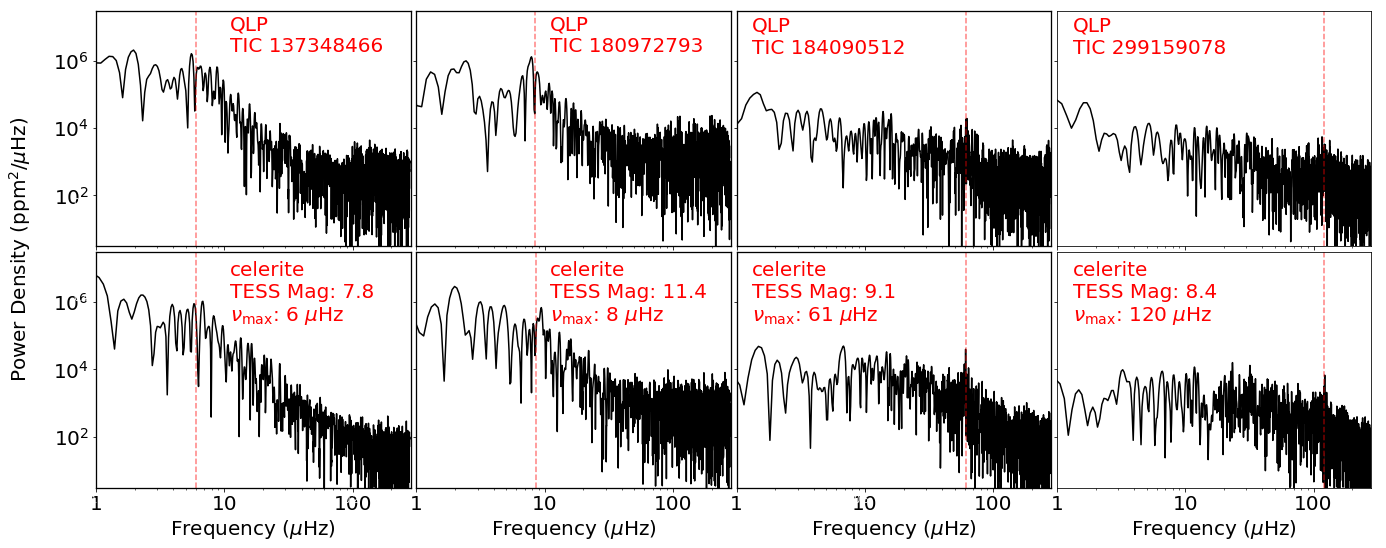}
    \caption{A comparison of real TESS power spectra (top row) with synthetic power spectra (bottom row) generated by \texttt{celerite} in this work. Each row corresponds to a red giant with a particular TESS magnitude that is oscillating at a specific $\nu_{\mathrm{max}}$.}
    \label{fig:appendix_simulations}
\end{figure}

Figure \ref{fig:appendix_simulations} shows example power spectra of oscillating red giants generated by \texttt{celerite} compared to real TESS giants with the same $\nu_{\mathrm{max}}$ and TESS magnitudes. There are differences between the noise levels of real and simulated data, which is typically attributed to the lack of noise contributions from both instrumental sources and flux contamination in simulated data. Additionally, in the absence of such contributions, the frequency-power profile of the simulated power spectra appear different to that of the real data, particularly at low frequencies. Future work will include more realistic simulations; however, the current data are sufficient to benchmark the network trained on \textit{Kepler}-as-TESS data in Section \ref{sec:threshold}). 

\section{Faint and Luminous Giants}
\label{Appendix:RG_hard}
\begin{figure}
    \centering
	\includegraphics[width=0.6\columnwidth]{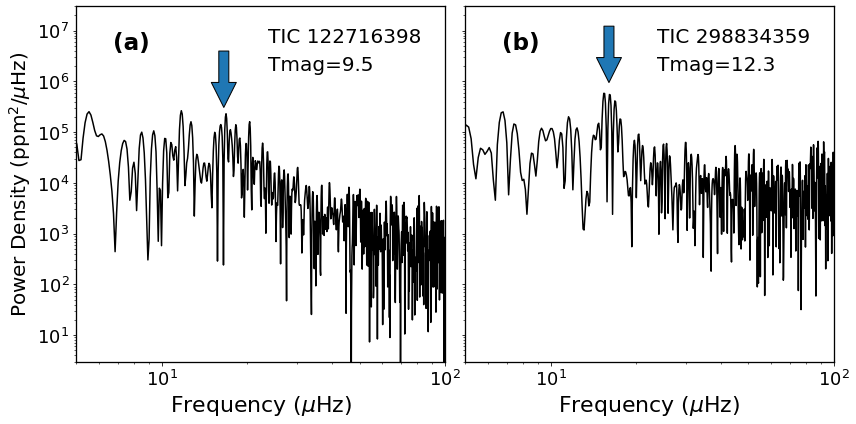}
    \caption{Power spectra for two luminous, oscillating red giants observed by TESS. The blue arrows indicate  $\nu_{\mathrm{max}}$, while $\mathrm{Tmag}$ indicates TESS magnitude. TIC 122716398 in panel (a) is a brighter and thus the `knee' of the granulation profile can be easily seen in the log-log power spectrum. TIC 298834359 in panel (b) is fainter and therefore its power spectrum has greater white noise levels. Oscillations for such faint and luminous giants typically appear as a sequence of peaks with heights marginally above a nearly flat frequency-power profile in the log-log spectrum.}
    \label{fig:appendix_faintgiants}
\end{figure}

Figure \ref{fig:appendix_faintgiants} shows power spectra of two oscillating red giants with $\nu_{\mathrm{max}}\sim15\mu$Hz. The giant in Figure \ref{fig:appendix_faintgiants}a is brighter than the giant in Figure \ref{fig:appendix_faintgiants}b and therefore has a power spectrum with lower noise levels. As a result, the slope of the granulation profile (red noise), which is an important feature for the neural network classifier in detecting low $\nu_{\mathrm{max}}$ oscillations \citep{Hon_2018}, can still be observed. In contrast, the power spectrum in Figure \ref{fig:appendix_faintgiants}b is more white noise-dominated, which obscures the granulation profile and makes the detection of oscillating luminous giants at fainter TESS magnitudes more difficult. 

\section{Examples of Oscillating Giants}
\label{appendix:sample_examples}
\begin{figure}
    \centering
	\includegraphics[width=\columnwidth]{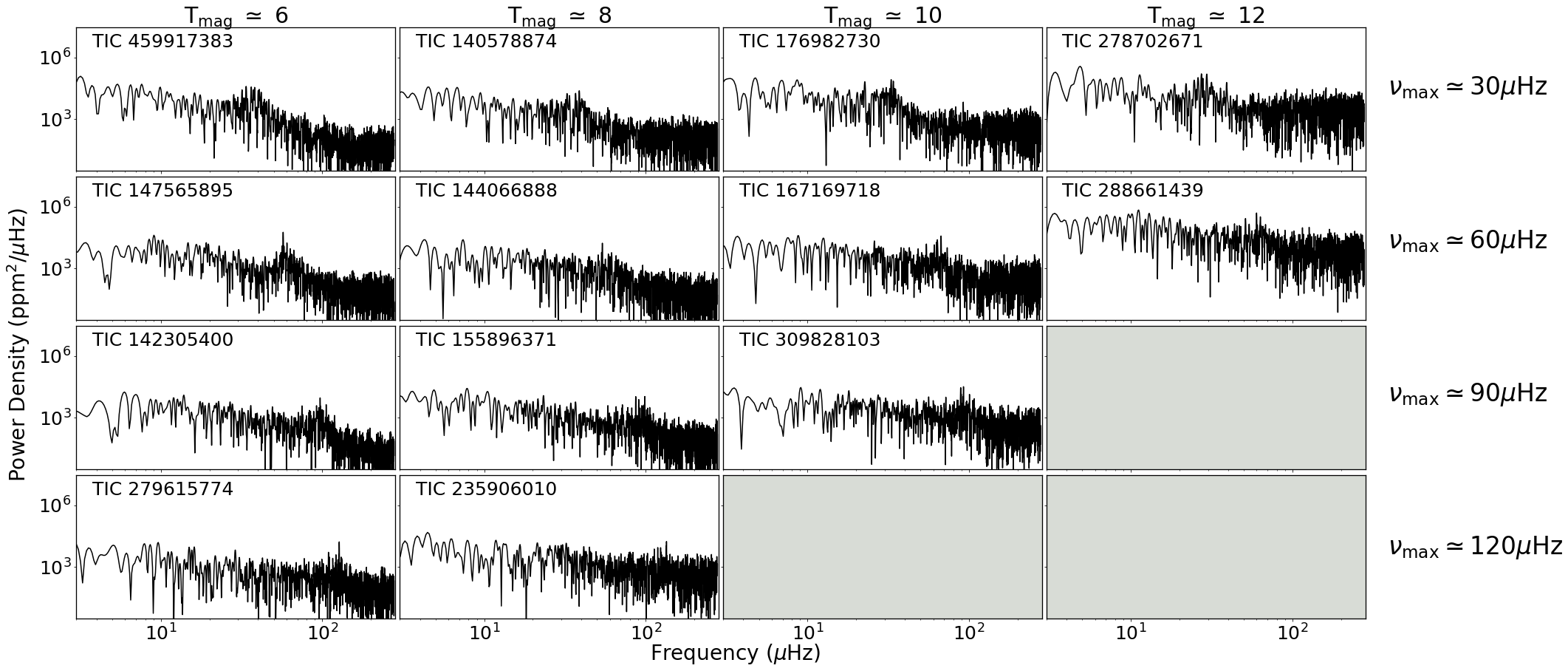}
    \caption{Examples of detected oscillating giants spanning $\nu_{\mathrm{max}}\sim30-120\mu$Hz. Each row corresponds to a particular $\nu_{\mathrm{max}}$ value, while each column corresponds to a specific TESS magnitude. The bottom right panels are greyed out because no oscillating giants are found with such values of $\nu_{\mathrm{max}}$ and TESS magnitudes in our sample as a consequence of the oscillation detection limit (Figure \ref{fig:results2}b).}
    \label{fig:appendix_lowlum}
\end{figure}

\begin{figure}
    \centering
	\includegraphics[width=\columnwidth]{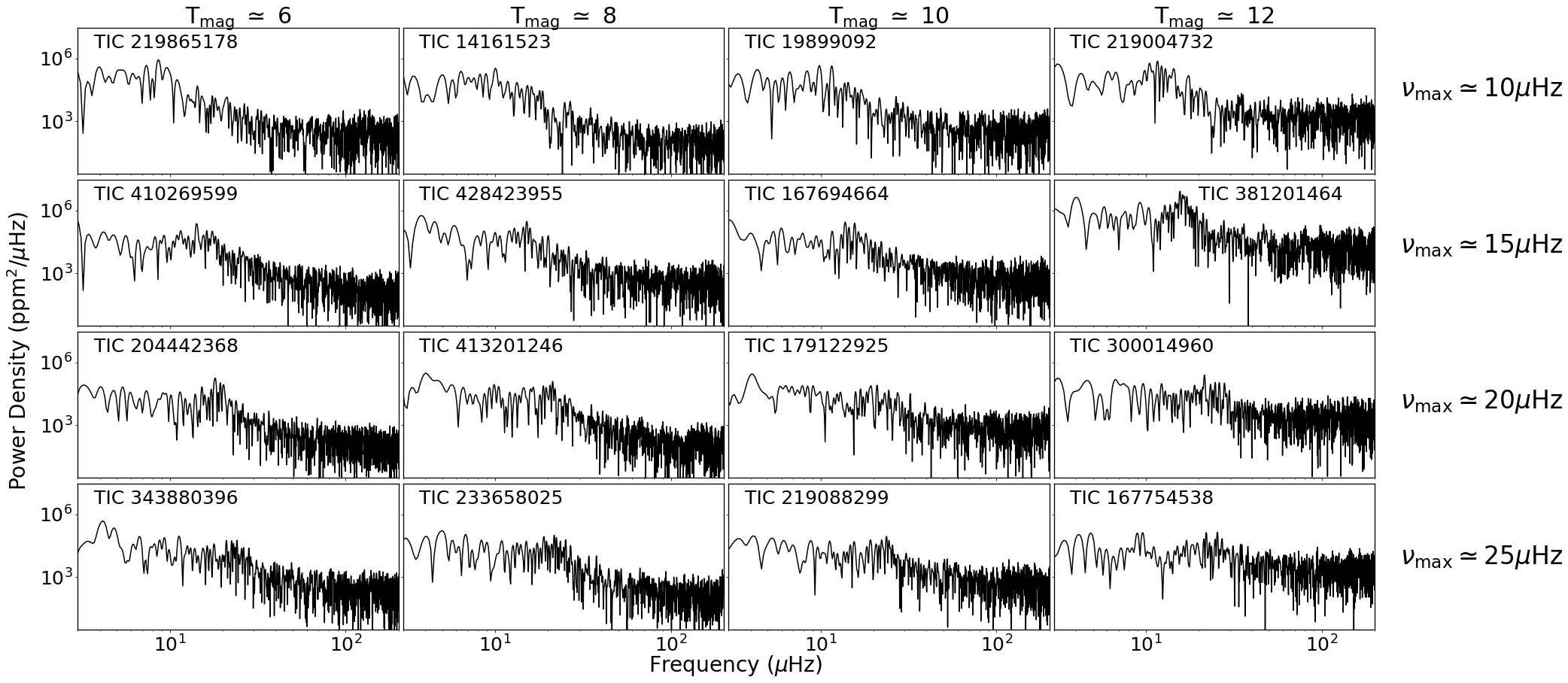}
    \caption{Same as Figure \ref{fig:appendix_lowlum}, but for oscillating giants spanning $\nu_{\mathrm{max}}\sim10-25\mu$Hz.}
    \label{fig:appendix_highlum}
\end{figure}

We show several examples of oscillating red giants spanning a higher range of $\nu_{\mathrm{max}}$ (lower luminosity) in Figure \ref{fig:appendix_lowlum}, while Figure \ref{fig:appendix_highlum} shows giants across a lower range of $\nu_{\mathrm{max}}$ (higher luminosity).

\bibliography{sample631}{}
\bibliographystyle{aasjournal}



\end{document}